\documentclass{article}
\usepackage{arxiv}
\usepackage[export]{adjustbox} 
\usepackage{placeins}
\usepackage{xcolor}
\usepackage[utf8]{inputenc} 
\usepackage[T1]{fontenc}    
\usepackage{hyperref}       
\usepackage{url}            
\usepackage{booktabs}       
\usepackage{amsfonts}       
\usepackage{amsmath}
\usepackage{nicefrac}       
\usepackage{microtype}    
\usepackage{lipsum}
\usepackage{graphicx}
  \usepackage[
    backend=biber,
    style=chem-acs,
    articletitle=true
  ]{biblatex}

\graphicspath{ {./images/} }

\title{Direct and Simultaneous Measurement of the Stiffness and Internal Friction of a Single Folded Protein}

\author{Surya Pratap S. Deopa, Shatruhan Singh Rajput, Aadarsh Kumar, Shivprasad Patil* \\
	Department of Physics\\
	Indian Institute of Science Education \& Research\\
	Pune 411008, Maharashtra, India \\
	\texttt{s.patil@iiserpune.ac.in} \\
	}

\begin{document}
\maketitle
\begin{abstract}
    The nanomechanical response of a folded single  protein,  the  natural nanomachine responsible for myriad  biological processes,  provides insight into its function. The conformational flexibility of a folded state, characterized by its viscoelasticity, allows proteins  to  adopt different shapes to perform  their function. Despite efforts, its direct measurement  has not been possible so far.  We present a direct and simultaneous measurement of the stiffness and internal friction of the folded domains of the protein titin using a special interferometer based atomic force microscope.   We analysed the data by carefully separating  different  contributions affecting the  response of the experimental  probe to obtain the folded state's  viscoelasticity.  Above $\sim$  95 pN of force, the individual immunoglobulins of titin   transition from an elastic solid-like native state to a soft viscoelastic intermediate.  
    
\end{abstract}
\begin{center}
        \begin{figure}[ht]
            \centering
            \textbf{TOC figure}\par\medskip
            \includegraphics[scale=1,center]{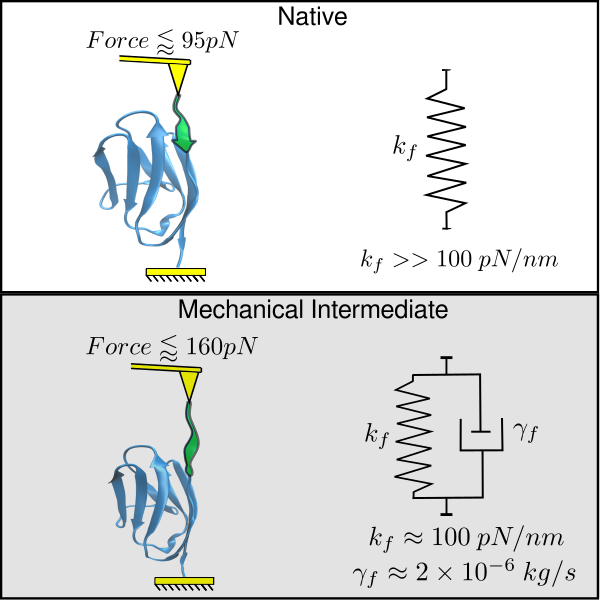}
        \end{figure}
    \end{center}   

Proteins are a type of heteropolymers,  which play a pivotal role in  biological processes responsible for life\autocite{Albe_2002_book,alberts1998cell,goodsell2009machinery}. They are natural nanomachines\autocite{Albe_2002_book}  evolved  to achieve specific tasks in cellular processes \autocite{Symmetry-Thiru, alberts1998cell}.   Proteins  need to be 
 dynamic\autocite{zaccai2000soft} and flexible\autocite{dong2009determination}. They also exhibit heterogeneity revealed by single molecule experiments\autocite{hyeon2014evidence,hinczewski2016directly}.  One of the  important goals of  nanotechnology research  is  to  quantify folded protein's  mechanical response, establish its connection with constituent  network of bonds and use this knowledge for design and fabrication of efficient nanomachines \autocite{soong2000powering,  howard2001mechanics}.
 Owing to  heterogeneity in folded states, it is important that  nanomechanical response  is measured at the level of single proteins\autocite{hyeon2014evidence}.    

From a nano-mechanics standpoint,  internal friction and stiffness  can be directly measured  by deforming a single protein at a certain strain rate.  To characterize the viscoelasticity of a single  protein one needs to strain it periodically and measure the in-phase and out-of-phase components of the  stress generated. Such measurements  provide both, stiffness and    internal friction coefficient  of the molecule.
 Although  viscoelastic behaviour in an ensemble of proteins has been reported\autocite{wang2010elasticity}, we lack a quantitative nanometrology tool that can  simultaneously measure stiffness and internal friction and  is able to provide  full characterization of viscoelasticity  of a single protein's folded state.

It is possible to deform a single folded protein and measure the resulting stress using AFM \autocite{dong2009determination}.  The conventional protein pulling experiments  measure  force required to unfold a single domain out of multiple domains  covalently connected to each other.  \autocite{carrion1999mechanical}.
In a typical saw-tooth pattern of such a  force-extension curve, the profile between two peaks is that of the entropic elasticity of unfolded chain which is around $\sim$ 10 pN/nm\autocite{rajput2020nano,khatri2008internal}. The folded domains are relatively stiffer ($\sim $200 pN/nm) \autocite{zaccai2000soft, dong2009determination} and hence are considered not to contribute to measured response. At a high stretch, where the folded domains stiffness is comparable to the unfolded chain, one is still faced with the challenge of separating them to obtain the response of the folded domains alone. It has been pointed out that pulling experiments using AFM  are not suitable to measure  viscoelasticity of the folded state\autocite{wang2011folded}.  

\begin{figure}
\includegraphics[width=\columnwidth]{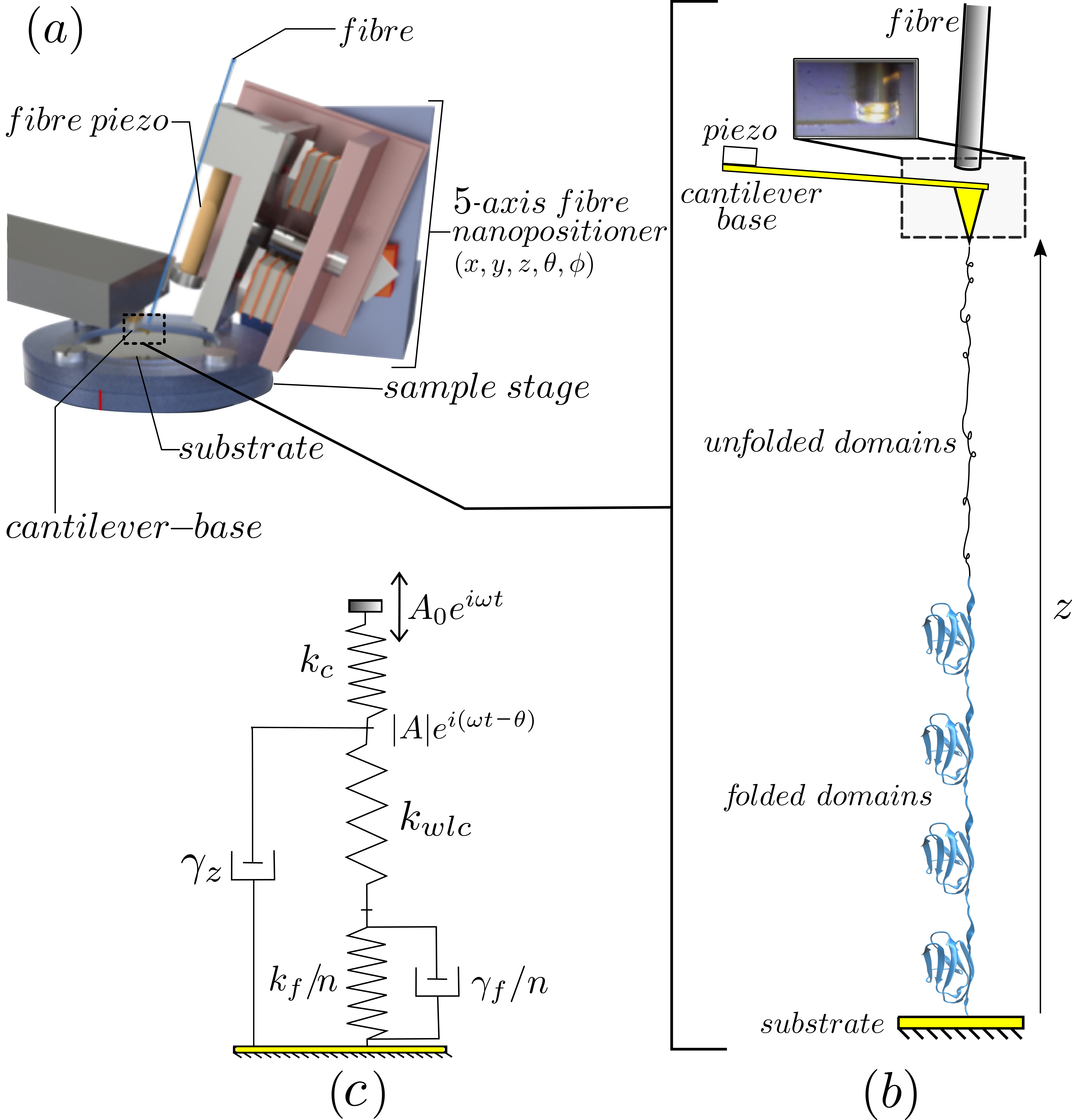}
\caption{\label{fig1} Measurement of folded domain's viscoelasticity using interferometer-based  AFM. a) Schematic of the  nano-positioner  used to place the fibre over the end of the cantilever beam.  The end of the fibre is coated with a semi-mirror and  it is  aligned parallel to the back surface of the cantilever using a five-axis inertial slider.    The amplitude of the molecular extension is directly measured as opposed to the conventional method in which cantilever bending is measured. 
b) The measurement scheme depicting the octamer tethered to the cantilever and substrate. The base is oscillated at off-resonance while the substrate is pulled away in a quasi-static manner to increase tip-sample separation z and the amplitude ($A$) and phase lag ($\theta$) are monitored. The inset shows the cleaved end of fiber coated with a semi-mirror aligned perpendicular to the length of the cantilever c) The cantilever-protein system is modelled with spring and dash-pots.   $A_0$ - base amplitude,  $A$ - tip amplitude, $\omega$- drive frequency, $\theta$- phase lag between  tip and  base of the cantilever, $k_c$ - cantilever stiffness  $\gamma_z$ hydrodynamic damping due to the proximity of the cantilever to the substrate,    $k_f$ and $\gamma_f$ -stiffness and  friction coefficient of folded domain respectively, $n$- number of folded domains. $k_{wlc}$- stiffness  of the  unfolded  Worm-Like-Chain (wlc).   Since $k_f/n$ and $k_{wlc}$ are in series, the weaker among the two springs has a dominant  contribution  to the amplitude $A$.}
\end{figure}

 \begin{figure}[ht]
\includegraphics[width=\columnwidth]{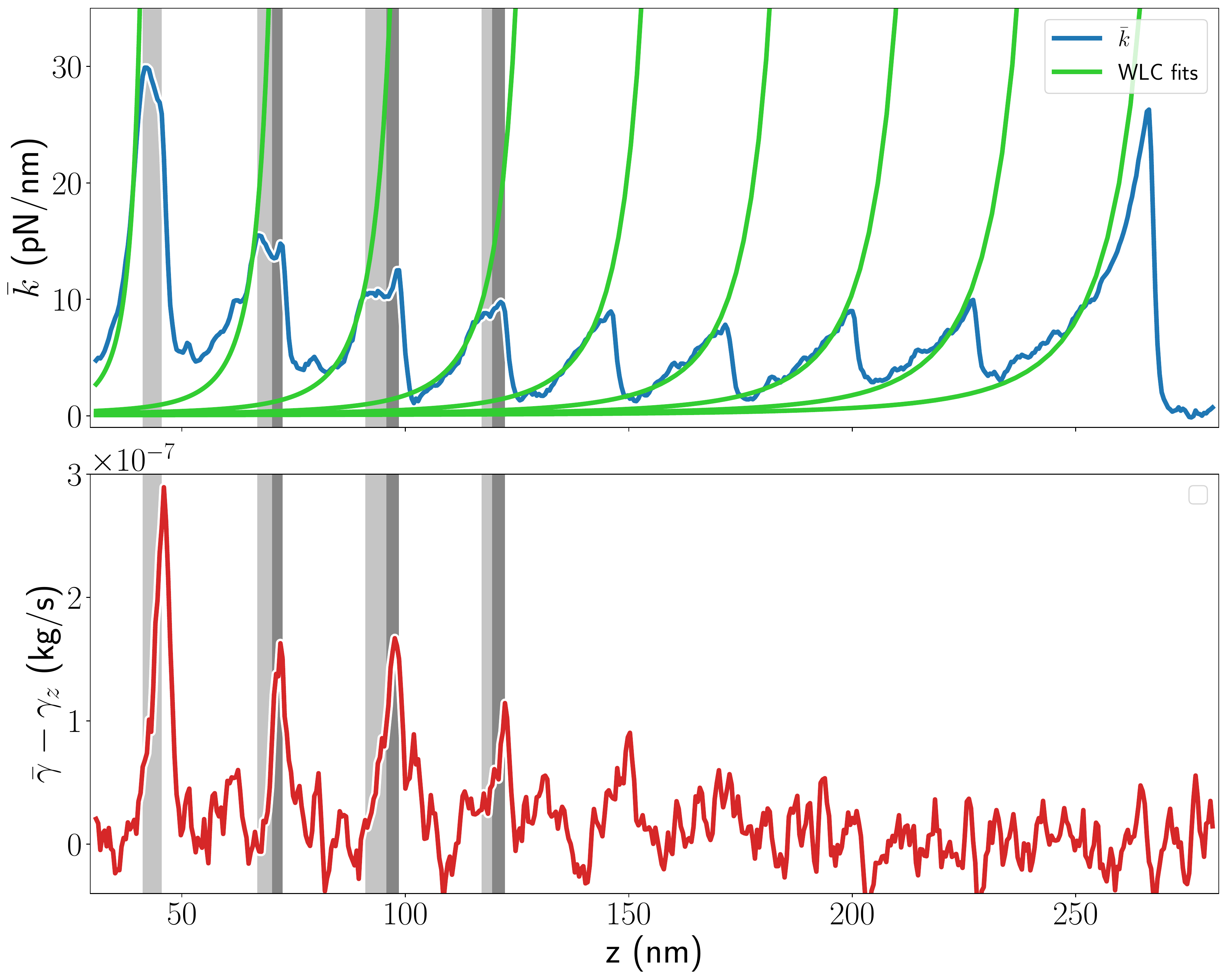}
\caption{\label{fig2}  
The measured combined stiffness (blue continuous) and friction  coefficient (red continuous) of  unfolded chain and the folded domains of I-27 octamer as domains  sequentially unfold. The amplitude and phase of the tip displacement is used for calculation. The data shows a sawtooth pattern of  unfolding events similar to constant velocity pulling experiments. The green continuous line is a fit using  wlc model. The mean difference in contour length between two consecutive peaks is 29 $\pm$  0.8 nm. A persistence length of 0.4 nm is used. The first four events deviate from wlc in the shaded region where folded domains are comparable in stiffness to the unfolded chain and contribute to the measurement. The dissipation is seen in the corresponding region due to the internal friction in folded domains. The data in the dark shaded region is used for further analysis to obtain the stiffness and internal friction coefficient of the folded state.    Cantilever stiffness is 0.6 N/m, The cantilever base  is excited  with  frequency 2.1 kHz and   amplitude  1 nm. The plot is representative of more than 50 traces recorded at different frequencies and amplitudes. The procedure to obtain the molecular friction coefficient  by separating the contribution from the cantilever damping at close proximity with the substrate is provided in supplementary materials, section 2.  }
\end{figure}

In this letter, we report  quantitative measurement of viscoelasticity of the folded state of a single protein using a novel interferometer based AFM.  By careful analysis,  we have separated the response of individual folded domains from  other contributions. The experiments demonstrate for the very first time that  direct and simultaneous measurement of internal friction  of folded protein is  possible.
The salient features of our instrument are i) it measures displacement of the cantilever and not  bending ii) allows use of stiff cantilevers ($\approx$ 0.6 N/m ), small amplitudes (0.1 to 1 nm) and  iii)in off-resonance condition,  performs artefact-free measurement of nano-scale viscoelasticity in liquid environments. See Fig. 1. Briefly, a fiber with  semi-mirror at its end is aligned perpendicular to a cantilever using  a five-axis fibre nanopositioner ( Fig.1 (a)).The mirror and back of the cantilever form a Fabry-Perot etalon. Infrared light(1310 nm) from a laser diode is guided into a 2 $\times$ 2  splitter.  50 $\%$  laser light goes to the etalon. The resulting interference  signal,  due to primary beam entering into the etalon and its  interference   with  multiple reflections between  mirror and  cantilever's back surface    is guided using the same splitter onto a photodiode. The current in the photodiode is extremely sensitive to fibre cantilever distance. We typically obtain sensitivity 300-500 mV/nm   and measure  cantilever displacement  with  precision of less than an angstrom.

We used repeats of 8 domains of Immunoglobulin (IgG) of titin (I-27)$_8$, which unfolds sequentially.
100 $\mu$l of I-27${_8}$ protein solution  with a concentration of 10mg/ml in PBS (pH 7.4) was drop cast  onto a  gold coated cover-slip  mounted in a fluid cell and excess protein is washed away after incubation.  The cantilever is lowered into the solution and fiber is positioned on the back of the cantilever above the tip. The semi-mirror on the fibre end is  aligned parallel to the back-surface of  cantilever. The cantilever base is oscillated  with amplitude $A_0$ at a frequency $\omega$,  which is far below its resonance. The amplitude $A$ and phase lag $\theta$ of the protein-attached tip is recorded using the interferometer,   as the mean separation  between the base and substrate($z$) is varied. The amplitude, $A$ and phase $\theta$ versus mean separation $z$  is used to calculate the stiffness and friction  coefficient profiles shown in Fig. 2, using  equations $\Bar{k} = k_c\Big(\frac{A_0}{|A|}\cos{\theta}-1\Big)$ and $\Bar{\gamma} = \frac{k_c A_0}{|A|\omega}\sin{\theta}$. See supplementary materials, section 1 and 2. 
At relatively large amplitudes  ($A_0\sim$1 nm), and high stretch (above $\sim$ 95 pN), a portion of the tip amplitude ($\sim$ 0.1 nm) is contributed  by the folded domains.    

  Figure 2 shows  measurement of  stiffness (continuous blue line) and friction  coefficient (continuous red line) of the polyprotein (I-27)$_8$ as the separation $z$ between the substrate and the cantilever base is increased in a quasi-static manner.   It is observed that for the first four unfolding domains, after the initial increase at lower extensions,  the stiffness abruptly goes down or plateaus and rises again( the shaded region in fig. 2). This is not seen in the last four domains. The total measured dissipation $\bar{\gamma}$  also decreases with  each consecutive unfolding peak.  Both observations are explained in Supplementary Information, section 3. ) 
  
The  Marko-Siggia approximation describes the force extension behaviour of unfolded domains in conventional experiments\autocite{marko1995stretching}. It provides a force-extension relationship, which can be fitted to  experimental force curves to obtain relevant parameters such as persistence length $p$ and  contour length $L_c$. The derivative of force with respect to extension provides a relationship between local stiffness $k_{wlc}$  and  end end-to-end molecular extension $z$ \autocite{marko1995stretching}.  
\begin{equation}
   k_{wlc} =  \frac{dF}{dz}=\frac{k_bT}{pL_c}\Big(\frac{1}{2(1-z/L_c)^3}+1\Big)
\end{equation} 

 At lower extensions in each unfolding profile  of first four domains, where  stiffness of  chain is much smaller compared to  folded domains, wlc model of entropic elasticity given by equation 1  fits well (green continuous line). In the shaded region, however, the measured stiffness (continuous blue line) is lower than  entropic stiffness . Corresponding to this region,  phase also shows lag and there is measurable dissipation of energy (continuous red line).

   We argue that in the  shaded region of Fig.\ref{fig2},  measured stiffness contains  contribution from folded domains.  We developed a method to separate it from  measured viscoelastic response in this region (See supplementary information, section 1 and 2.) 
   The conventional  constant velocity pulling experiments using AFM have measured   mechanical stability of I-27 domains. It mechanically unfolds via an intermediate. The domains fall into mechanical  intermediate above $\sim$95 pN and completely unfold at $\sim$ 200 pN\autocite{marszalek1999mechanical}. See Fig. 3. 
   Note that when the response deviates from  wlc description in Fig.2, a folded domain is  likely to be in either  native or  intermediate state and hence  number of domains in either state is not known. However, as $z$ is  increased, loading the folded domains further, all  of them fall into  intermediate state, each adding 0.66 nm of contour length to the chain\autocite{marszalek1999mechanical,nunes2010force,taniguchi2008effect}. At this point, as seen in figure 2, the stiffness starts to rise again before the next domain is unfolded. In this region, shaded in dark, we use our analysis method to obtain the stiffness and friction  coefficient of the folded domains.      
   
 \begin{figure}[ht]
\centering
\includegraphics[width=0.75\columnwidth]{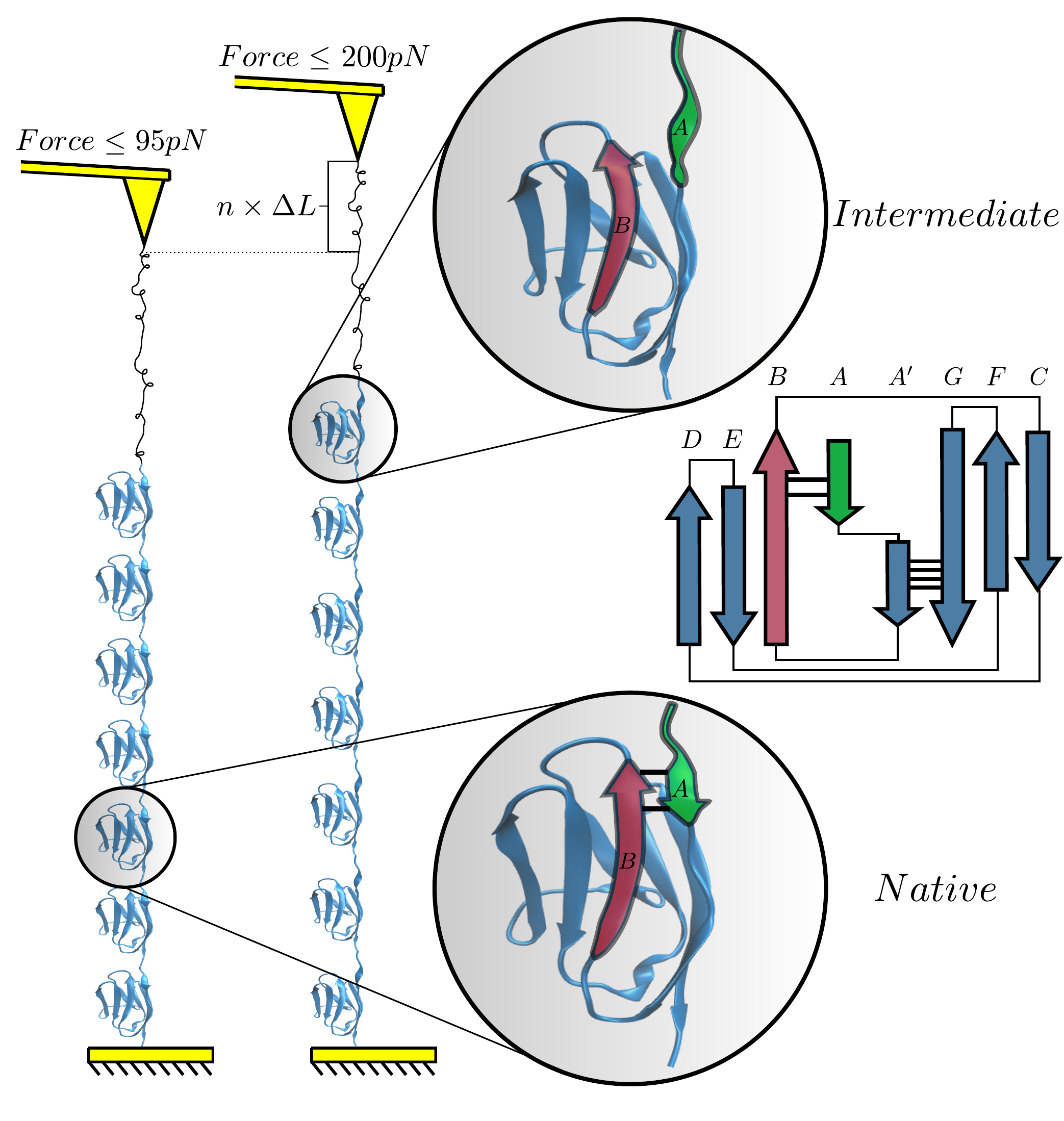}
\caption{\label{fig3} The octomer of I-27 under force. The cartoon shows a situation, wherein one  domain is unfolded and the remaining seven are in the folded state.  When the folded domains are pulled on by increasing $z$, the force is applied to  domains through  wlc chain of   first unfolded domain. When  force exceeds  $\approx$ 95 pN ( shaded region in fig. 2), the domains  in the native state (left) are pushed  into  mechanical intermediate(right). As shown in the $\beta$-sheet schematic, the hydrogen bond network between B-A and A'-G is responsible for the  mechanical stability of the native state, of  which the network between A and B is broken for the intermediate, elongating the protein by $n \times \Delta L $, where $\Delta L$ =  0.66 nm \autocite{marszalek1999mechanical}. When all the domains enter into the intermediate, the stiffness starts to rise again as shown in the darker shaded area in figure 2. Measurements in this region are used to estimate the stiffness and friction coefficient of the individual folded intermediates.       
}
\end{figure}

\begin{figure}[ht]
\includegraphics[width=\columnwidth]{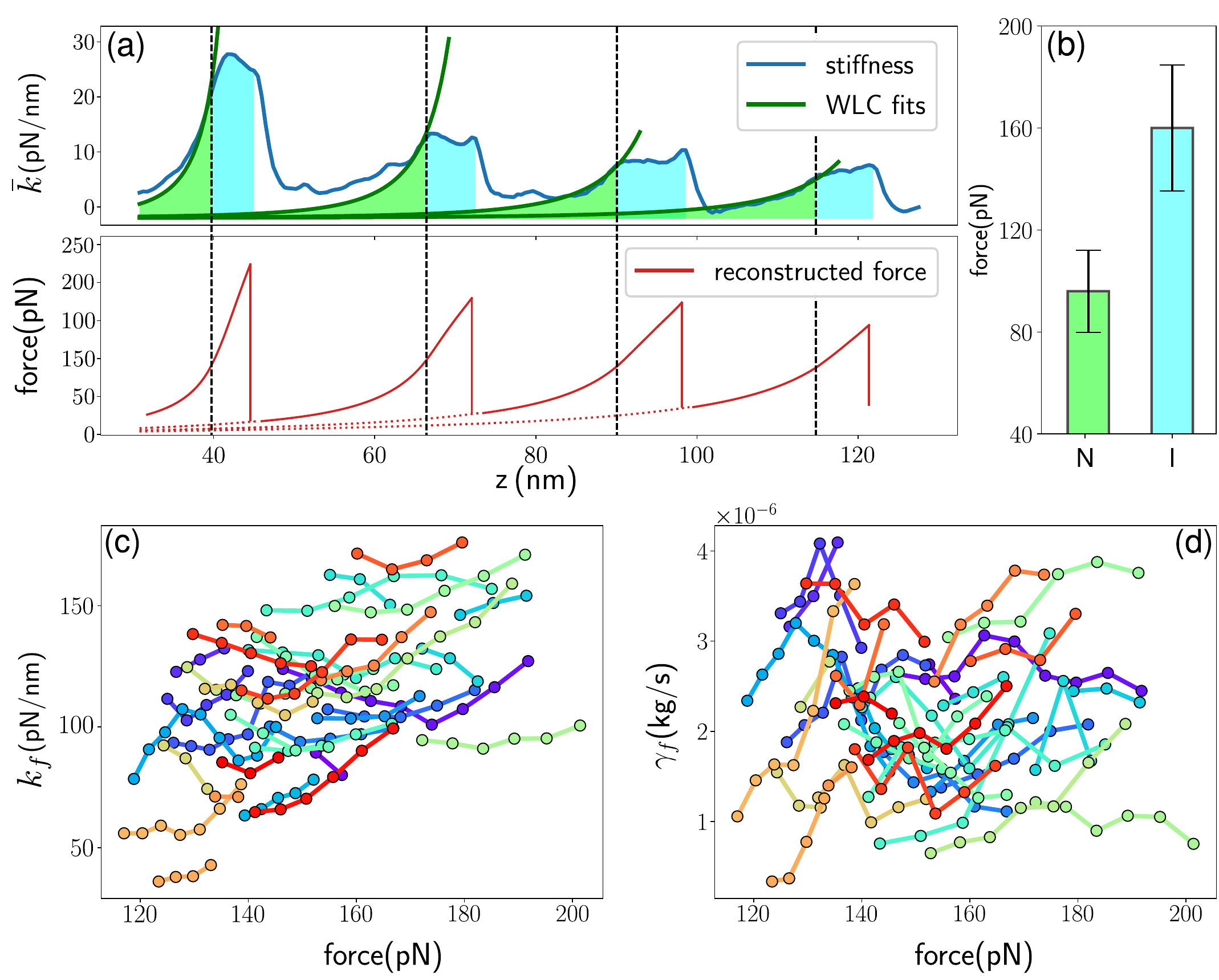}
\caption{\label{fig4} a)The integration of stiffness-extension (continuous blue line) provides  force-extension plot ( continuous red line). The vertical dotted line indicates  deviation of stiffness-extension from wlc model fit (continuous green lines) of equation 1. Its intersection with force-extension gives force at which,  stiffness has a contribution from folded domains. b) Bar plots of force where  transition occurs and the force required to unfold a domain completely (N: Native and I: Intermediate). The deviation kicks-in when force exceeds  95 $\pm$ 16 pN.  The force required to unfold the domain completely is 161 $\pm$ 25 pN. These values match with force needed to push domains of (I-27)$_8$ into the intermediate and then unfold it completely\autocite{marszalek1999mechanical}. The mean force required to unfold  a domain completely, is $\sim$ 200 pN in static pulling experiments. It is $\sim$ 160 pN in our experiments due to the energy provided to it through oscillations.      
c) stiffness $k_f$ and friction  coefficient $\gamma_f$ of a single folded domain at different forces $F$.  The plots   provide  viscoelastic characterization of single folded intermediate. The data points in the same colour are from the analysis of   same  unfolding peak. The data is pulled together from many different  profiles similar to the one shown in Fig. 2.}  
\end{figure}
   
   In our experiments, the oscillating base of the cantilever and the substrate are  pulled away from each other  with protein octomer I-27${_8}$ tethered between the cantilever tip and the substrate (Fig.1).  The measured response of the tip ( amplitude $A$ and phase $\theta$ )  is due to i)that of  the individual folded domains, ii) worm-like chain of unfolded domains,  iii) freely jointed chain formed of folded domains.
   However, the freely jointed chain is 95\% extended and is much stiffer compared to the first two at forces above 20 pN. Hence it does not contribute to the modulation of tip amplitude and only adds a constant length to the unfolded chain. We need to separate the contribution from wlc chain of unfolded domains  to the measured stiffness in order to single out  the viscoelastic response of  folded domains. We discuss the analysis procedure followed to obtain stiffness and friction coefficient  of individual folded domains.  
   For a given $z$, the force balancing equation for the spring and dash-pot arrangement of  Fig.1 gives us 
\begin{equation}
    k_c(A_0-A)= \Bigg[\Big(\frac{n}{k_{f} + i\gamma_f \omega} + \frac{1}{k_{wlc}}\Big)^{-1} + i\gamma_z\omega\Bigg]A
\end{equation}
Where $k_{wlc}$ is chain stiffness and $k_f$ and $\gamma_f$ are stiffness and friction  coefficient of the folded domain respectively. $\gamma_z$ is damping  provided by the surface to the  cantilever body. 
We get expression for stiffness and friction coefficient of individual folded monomers as ( See Supplementary Information, section 1 and 2 for details of this calculation) 
\begin{equation}
       k_f=\dfrac{n\Big(\frac{\bar{k}}{\bar{k}^2 + \omega^2 \gamma^2} - \frac{1}{k_{wlc}}\Big)}{\Big(\frac{\bar{k}}{\bar{k}^2 + \omega^2 \gamma^2} - \frac{1}{k_{wlc}}\Big)^2 + \Big(\frac{\gamma\omega}{\bar{k}^2 + \omega^2 \gamma^2}\Big)^2} 
\end{equation}
\begin{equation}
       \gamma_f=\dfrac{n\Big(\frac{\gamma\omega}{\bar{k}^2 + \omega^2 \gamma^2}\Big)}{\Big(\frac{\bar{k}}{\bar{k}^2 + \omega^2\gamma^2} - \frac{1}{k_{wlc}}\Big)^2 + \Big(\frac{\gamma\omega}{\bar{k}^2 + \omega^2 \gamma^2}\Big)^2} 
\end{equation}
where $\bar{k}$ and $\bar{\gamma}$ are measured  stiffness and friction  coefficient,  $\gamma=\bar{\gamma}-\gamma_z$. We need an estimate of  $k_{wlc}$ to use equations 3 and 4 in order to get  viscoelasticity of a folded  domain from measured $\Bar{k}$ and $\Bar{\gamma}$. For this,  we integrated the measured "$\Bar{k} -z $" curve in Fig. \ref{fig2} and obtained "Force-z" curve(Fig.3). In the grey shaded region of Fig.2, this force produces extensions both in  wlc chain and the folded domains. The $k_{wlc}- F $ relationship in high force limit ($F>>\frac{k_BT}{p}$) where the polymer is significantly extended is given by \autocite{khatri2008internal}- $k_{wlc}(F)=\frac{4}{L_c}\sqrt{\frac{p}{k_BT}}F^{\frac{3}{2}}$.   The $k_{wlc}$ is then estimated since the force on the chain for each $z$ is  determined.  Equations 3 and 4 are used to calculate  stiffness and friction  coefficient of individual intermediates from  experimental data.
   
 The numerical integration of   stiffness-extension data,   presented in Fig.\ref{fig2},  provides  force-extension curve as shown Fig.4a. We use only first four domains from Fig. 2 for this analysis. Such force-extension profiles inform  forces at which  folded domains start to contribute to  stiffness measurement. This is shown by dotted vertical lines intersecting the red curve depicting the force profile. The force at this intersection is threshold force after which,   folded domain's stiffness contributes to  measurement. This occurs at 95$\pm$ 16 pN. The protein  unfolds completely  at 161 $\pm$ 25 pN.  Fig. 4 b shows bar plots obtained by this procedure for all our experimental data. Using  dark shaded region in Fig. 2  for  our analysis,   we computed  stiffness ($k_f$) and friction coefficient ($\gamma_f$) of individual folded domains at different forces   shown in    Fig. 4c and d.     
 
 Fig. 4b  indicates that folded protein's response starts to contribute to   stiffness measurement when force on it exceeds $\sim$ 95 pN.  This matches  well with  force required to   make the transition from native to mechanical intermediate  \autocite{marszalek1999mechanical,taniguchi2008effect,nunes2010force}.  It suggests that our measurement of viscoelasticity is that of  the folded intermediate, which is soft. The domains remain in the native state below $\approx$ 95 pN. In this region, which is not shaded in gray in Fig. 2a, the data fits  wlc behavior. It indicates that in this region  folded domains in the native state are too stiff to make any discernible effect on tip amplitude. It is fully elastic having immeasurably high stiffness compared to the unfolded chain. Under the loading force F, the mechanical intermediate's stiffness increases slightly with force, whereas the friction  coefficient does not show any appreciable change. 
 
 It has not been possible to measure viscoelastic response of single folded protein using AFM, a technique at the forefront of pulling single proteins. The conventional deflection detection  scheme,  used to measure  cantilever response, records   cantilever bending as opposed to displacement in our experiments. (See Supplementary Information, section 4). It is important to operate off-resonance in order to avoid the stiffness change in the molecule affecting the phase, which produces measurement artefacts. The bending signal measured with deflection detection scheme  is immeasurably low in off-resonance operation ( See Supplementary Information, Fig. S12.)  

It is possible to estimate Young's modulus $G$ and the viscosity $\eta$ from the measurement of stiffness and friction coefficient using dimensional analysis argument. $G =  k \times L/A$, Where L is length, A is the cross sectional area of the object and $k$ is its stiffness.  Since we do not know the orientation of protein, taking it to be a cube of side 4 nm, results in an estimate of $G \sim$ 0.2 GPa. Note that the native state  has an even higher value of $G$. $G_{glass}$  is $\sim$ 50 GPa and $G_{steel}$ is $\sim$ 200 GPa.  
 For the estimate of $\eta$, we take an object of $\sqrt{bd} $ moving in a medium of viscosity $\eta$. $d$ is extension in the folded domain.  $\gamma = 6\pi \eta \sqrt{bd} $; $\eta \sim $ 150 Pa.s. This is in the same order with earlier estimates by group of Hansma from tip relaxation on Lysozyme\autocite{radmacher1994imaging}, however  a factor of $\sim 10^5$ times higher than the ones estimated from folding-refolding experiments \autocite{Cellmer-Eaton-2008internal-friction},  which Wang and Zocchi attributed to shear thinning, $\eta \sim 1/\omega^2$ \autocite{wang2011folded}.
 It is likely that,  using the folding rate dependence on  solvent viscosity, one measures  internal friction of the transition state, rather than that of the folded structures, as in this work.

 Considering the folded intermediate as a viscoelastic element, its relaxation time   is $\tau=\gamma_f/k_f \approx 10 $ $\mu$s. The dissipation of energy by  intermediate at $\mu$s timescales suggests that the oscillatory perturbation provided to  reaction co-ordinate couples with a slowly relaxing mode.  The all-atom simulations  have shown that structural relaxation processes involving the protein backbone are in the microsecond time scale\autocite{shaw2010atomic}.  The high stiffness and lack of dissipation of the native state suggest that, in this case, modes coupled to the reaction coordinate relax much faster. This drastic change in dynamics by merely breaking the hydrogen bond network between strands A and B,  keeping the structure otherwise intact, is an  intriguing new aspect of protein dynamics.   
 
 It has been shown in pulling experiments that  transition from native to mechanical intermediate of I-27 does not depend on temperature or pulling speed, whereas the unfolding force of the intermediate shows this dependence clearly\autocite{taniguchi2008effect,nunes2010force}. Based on this, it has been argued that  energy landscape of the intermediate is rugged whereas the native  state of I-27  works as a force buffer to protect it from large physiological forces\autocite{nunes2010force} . Our measurements are  consistent with these findings. We do not see any dissipation in the folded native state and  stiffness drops from an immeasurably high value to $\sim$ 100 pN/nm as the protein transitions from native to  mechanical intermediate accompanied by dissipation, indicating ruggedness.
 
 Borrowed from the theory describing amorphous glassy systems\autocite{frauenfelder1991energy}, the energy landscape description used in  the study of protein's folding-refolding dynamics is empirically successful, at least for small proteins.  However, this community debates the question -  to what extent  the projection of multi-dimensional   landscape onto a single reaction co-ordinate is accurate\autocite{neupane2016protein}. The success of one dimensional picture relies on the separation of fluctuation time-scales of the chosen co-ordinate with the rest\autocite{neupane2016protein}.We have provided the first direct observation of internal friction in a folded state through stress-strain analysis. Our measurement scheme  provides a direct test of such coupling between the end-to-end distance co-ordinate and other dimensions in the reduced,  multidimensional protein landscape. The  internal friction
measured here is a result of energy loss in  driving the folded state through many of its substates and it may provide useful insights into the inter-conversion  rates\autocite{horowitz2017minimum}.  The observed dissipation in our experiments is not possible unless the end-to-end distance reaction coordinate, which is modulated by the tip at $\sim$ 2 KHz is coupled to slowly relaxing backbone dynamics\autocite{shaw2010atomic}. The forced oscillations of end-to-end distance at $\sim ms$ timescale results in dissipation through such coupling. The data on lower operational frequency is presented in Supplementary Information, section 5.

 The protein of choice in our experiments is not only a paradigm in pulling experiments,  but also provides us a  native elastic state and a viscoelastic mechanical intermediate in one experiment. We show here that only the network of  hydrogen bonds between $\beta$ strands A and B decide the fate of mechanics governing IgG domains. The domain behaves like a Hookean solid,  having  no dissipation with the network intact,  whereas it has viscous elements,  if the network is broken. Understanding  effects of such local changes on global properties of folded macromolecules  are crucial for  unraveling the mechanism of allostery\autocite{Schachman,weber1972ligand} and could provide an instructive design principle for  synthesizing  nanomachines using molecular dynamics simulations\autocite{soong2000powering}.    
  
 While our stiffness measurements  match well with previous efforts\autocite{dong2009determination, zaccai2000soft}, the   friction coefficient of single protein's  folded state  are not yet measured. We discuss new possibilities opened with this  ability.
 Internal friction implies a rough energy landscape and slow dynamics due to frustration.
 Our direct measurement of dissipation  also reveals that  intermediate state is  frustrated.  It  allows estimate of effective   diffusion coefficient through  Stoke-Einstein relationship $D = K_BT/\gamma$, where $\gamma$ is the friction  coefficient. From our measurement of $\gamma$ for the intermediate, the diffusion coefficient is   $D = (1.94 \pm 0.05) \times$ 10$^{3}$ nm$^2/$s. It is possible to estimate ruggedness ($\epsilon$) by  using  Zwanzig relationship and measuring $D$ at two different temperatures\autocite{zwanzig1988diffusion}.  
 Secondly, using our experimental scheme , where the folded state is driven out of equilibrium,  the rupture-forces distribution along with energy dissipation to maintain the non-equilibrium state  can be used to extract information about the dynamics of slow processes\autocite{horowitz2017minimum}, which are poorly understood at present. 
 
   In conclusion, we simultaneously and directly measure stiffness and  internal friction  of single folded protein domain for the first time. In particular, we find  that beyond  a threshold force,  the IgG domain of titin moves  from an elastic solid-like native state to a viscoelastic mechanical intermediate, which dissipates energy, providing the first direct evidence of internal friction.    The capability demonstrated in our experiment   provides hope of relating  mechanics to chemical details at the level of single bonds by coupling this type of measurement to suitable single-molecule optical spectroscopy, such as F{\"o}rster Resonance Energy Transfer. 
  \subsection*{Supplementary Description}
  Model: Protein-Cantilever system; Data Analysis; Features observed in stiffness vs extension and friction vs extension profiles; Fibre Interferometer AFM and single molecule experiments; Measurement at lower frequency.

  \subsection*{Acknowledgement}
   SPD acknowledges fellowship from IISER Pune. The work is supported by Wellcome-Trust DBT India alliance through intermediate fellowship to SP (500172/Z/09/Z).   

\printbibliography  

@book{Albe_2002_book,
  author = {Alberts, B. and Bray, D. and Lewis, J. and Raff, M. and Roberts, K. and Watson, J.D.},
  edition = {4th},
  publisher = {Garland, New York},
  title = {{Molecular Biology of the Cell}},
  year = 2002
}

@article{alberts1998cell,
  title={The cell as a collection overview of protein machines: Preparing the next generation of molecular biologists},
  author={Alberts, Bruce},
  journal={cell},
  volume={92},
  pages={291--294},
  year={1998}
}

@book{goodsell2009machinery,
  title={The machinery of life},
  author={Goodsell, David S},
  year={2009},
  publisher={Copernicus New York, NY}
}

@book{howard2001mechanics,
  title={Mechanics of motor proteins and the cytoskeleton},
  author={Howard, Jonathon},
  year={2001},
  publisher={Sinauer associates Sunderland, MA}
}

@article{zaccai2000soft,
  title={How soft is a protein? A protein dynamics force constant measured by neutron scattering},
  author={Zaccai, Giuseppe},
  journal={Science},
  volume={288},
  number={5471},
  pages={1604--1607},
  year={2000},
  publisher={American Association for the Advancement of Science}
}

@article{frauenfelder1991energy,
  title={The energy landscapes and motions of proteins},
  author={Frauenfelder, Hans and Sligar, Stephen G and Wolynes, Peter G},
  journal={Science},
  volume={254},
  number={5038},
  pages={1598--1603},
  year={1991},
  publisher={American Association for the Advancement of Science}
}

@article{dong2009determination,
  title={Determination of protein structural flexibility by microsecond force spectroscopy},
  author={Dong, Mingdong and Husale, Sudhir and Sahin, Ozgur},
  journal={Nature nanotechnology},
  volume={4},
  number={8},
  pages={514--517},
  year={2009},
  publisher={Nature Publishing Group}
}

@article{zwanzig1988diffusion,
  title={Diffusion in a rough potential},
  author={Zwanzig, Robert},
  journal={Proceedings of the National Academy of Sciences},
  volume={85},
  number={7},
  pages={2029--2030},
  year={1988},
  publisher={National Acad Sciences}
}

@article{carrion1999mechanical,
  title={Mechanical and chemical unfolding of a single protein: a comparison},
  author={Carrion-Vazquez, Mariano and Oberhauser, Andres F and Fowler, Susan B and Marszalek, Piotr E and Broedel, Sheldon E and Clarke, Jane and Fernandez, Julio M},
  journal={Proceedings of the National Academy of Sciences},
  volume={96},
  number={7},
  pages={3694--3699},
  year={1999},
  publisher={National Acad Sciences}
}

@article{kiracofe2011quantitative,
  title={Quantitative force and dissipation measurements in liquids using piezo-excited atomic force microscopy: a unifying theory},
  author={Kiracofe, Daniel and Raman, Arvind},
  journal={Nanotechnology},
  volume={22},
  number={48},
  pages={485502},
  year={2011},
  publisher={IOP Publishing}
}

@article{o2006comment,
  title={Comment on “oscillatory dissipation of a simple confined liquid”},
  author={O’Shea, SJ},
  journal={Physical review letters},
  volume={97},
  number={17},
  pages={179601},
  year={2006},
  publisher={APS}
}

@article{hyeon2014evidence,
  title={Evidence of disorder in biological molecules from single molecule pulling experiments},
  author={Hyeon, Changbong and Hinczewski, Michael and Thirumalai, D},
  journal={Physical review letters},
  volume={112},
  number={13},
  pages={138101},
  year={2014},
  publisher={APS}
}

@article{rajput2020nano,
  title={The nano-scale viscoelasticity using atomic force microscopy in liquid environment},
  author={Rajput, Shatruhan Singh and Deopa, Surya Pratap S and Yadav, Jyoti and Ahlawat, Vikhyaat and Talele, Saurabh and Patil, Shivprasad},
  journal={Nanotechnology},
  volume={32},
  number={8},
  pages={085103},
  year={2020},
  publisher={IOP Publishing}
}

@article{wang2010elasticity,
  title={Elasticity of globular proteins measured from the ac susceptibility},
  author={Wang, Yong and Zocchi, Giovanni},
  journal={Physical review letters},
  volume={105},
  number={23},
  pages={238104},
  year={2010},
  publisher={APS}
}

@article{wang2011folded,
  title={The folded protein as a viscoelastic solid},
  author={Wang, Yong and Zocchi, Giovanni},
  journal={EPL (Europhysics Letters)},
  volume={96},
  number={1},
  pages={18003},
  year={2011},
  publisher={IOP Publishing}
}

@article{janovjak2005molecular,
  title={Molecular force modulation spectroscopy revealing the dynamic response of single bacteriorhodopsins},
  author={Janovjak, Harald and M{\"u}ller, Daniel J and Humphris, Andrew DL},
  journal={Biophysical journal},
  volume={88},
  number={2},
  pages={1423--1431},
  year={2005},
  publisher={Elsevier}
}

@article{taniguchi2010dynamics,
  title={Dynamics of the coiled-coil unfolding transition of myosin rod probed by dissipation force spectrum},
  author={Taniguchi, Yukinori and Khatri, Bhavin S and Brockwell, David J and Paci, Emanuele and Kawakami, Masaru},
  journal={Biophysical journal},
  volume={99},
  number={1},
  pages={257--262},
  year={2010},
  publisher={Elsevier}
}

@article{nunes2010force,
  title={A “force buffer” protecting immunoglobulin titin},
  author={Nunes, Jo{\~a}o M and Hensen, Ulf and Ge, Lin and Lipinsky, Manuela and Helenius, Jonne and Grubm{\"u}ller, Helmut and Muller, Daniel J},
  journal={Angewandte Chemie International Edition},
  volume={49},
  number={20},
  pages={3528--3531},
  year={2010},
  publisher={Wiley Online Library}
}

@article{taniguchi2008effect,
  title={The effect of temperature on mechanical resistance of the native and intermediate states of I27},
  author={Taniguchi, Yukinori and Brockwell, David J and Kawakami, Masaru},
  journal={Biophysical journal},
  volume={95},
  number={11},
  pages={5296--5305},
  year={2008},
  publisher={Elsevier}
}

@article{radmacher1994imaging,
  title={Imaging adhesion forces and elasticity of lysozyme adsorbed on mica with the atomic force microscope},
  author={Radmacher, Manfred and Fritz, Monika and Cleveland, Jason P and Walters, Deron A and Hansma, Paul K},
  journal={Langmuir},
  volume={10},
  number={10},
  pages={3809--3814},
  year={1994},
  publisher={ACS Publications}
}

@article{hinczewski2016directly,
  title={Directly measuring single-molecule heterogeneity using force spectroscopy},
  author={Hinczewski, Michael and Hyeon, Changbong and Thirumalai, Devarajan},
  journal={Proceedings of the National Academy of Sciences},
  volume={113},
  number={27},
  pages={E3852--E3861},
  year={2016},
  publisher={National Acad Sciences}
}

@article{shaw2010atomic,
  title={Atomic-level characterization of the structural dynamics of proteins},
  author={Shaw, David E and Maragakis, Paul and Lindorff-Larsen, Kresten and Piana, Stefano and Dror, Ron O and Eastwood, Michael P and Bank, Joseph A and Jumper, John M and Salmon, John K and Shan, Yibing and others},
  journal={Science},
  volume={330},
  number={6002},
  pages={341--346},
  year={2010},
  publisher={American Association for the Advancement of Science}
}

@article{benedetti2016can,
  title={Can dissipative properties of single molecules be extracted from a force spectroscopy experiment?},
  author={Benedetti, Fabrizio and Gazizova, Yulia and Kulik, Andrzej J and Marszalek, Piotr E and Klinov, Dmitry V and Dietler, Giovanni and Sekatskii, Sergey K},
  journal={Biophysical journal},
  volume={111},
  number={6},
  pages={1163--1172},
  year={2016},
  publisher={Elsevier}
}

@article{Cellmer-Eaton-2008internal-friction,
  title={Measuring internal friction of an ultrafast-folding protein},
  author={Tseng, Chiao-Yu and Wang, Andrew and Zocchi, Giovanni and Rolih, Biljana and Levine, Alex J
   Cellmer,Troy and   Henry, E. R.  and Hofrichter, J and  Eaton, W. A.  
  },
  journal={Proceedings of National  Academy of Sciences},
  volume={105},
  number={47},
  pages={18320},
  year={2008},
  publisher={ National Academy of Sceinces }
}

@article{khatri2008internal,
  title={Internal friction of single polypeptide chains at high stretch},
  author={Khatri, Bhavin S and Byrne, Katherine and Kawakami, Masaru and Brockwell, David J and Smith, D Alastair and Radford, Sheena E and McLeish, Tom CB},
  journal={Faraday discussions},
  volume={139},
  pages={35--51},
  year={2008},
  publisher={Royal Society of Chemistry}
}

@article{Schachman,
  title={Still looking for ivory tower},
  author={Schachman, H Howard K},
  journal = {Annual Review of Biochemistry},
  volume={69},
  number={},
  pages={1 --69},
  year={2000},
  publisher={ACS Publications}
}

@article{neupane2016protein,
  title={Protein folding trajectories can be described quantitatively by one-dimensional diffusion over measured energy landscapes},
  author={Neupane, Krishna and Manuel, Ajay P and Woodside, Michael T},
  journal={Nature Physics},
  volume={12},
  number={7},
  pages={700--703},
  year={2016},
  publisher={Nature Publishing Group}
}

@article{soong2000powering,
  title={Powering an inorganic nanodevice with a biomolecular motor},
  author={Soong, Ricky K and Bachand, George D and Neves, Hercules P and Olkhovets, Anatoli G and Craighead, Harold G and Montemagno, Carlo D},
  journal={Science},
  volume={290},
  number={5496},
  pages={1555--1558},
  year={2000},
  publisher={American Association for the Advancement of Science}
}

@article{weber1972ligand,
  title={Ligand binding and internal equilibiums in proteins},
  author={Weber, Gregorio},
  journal={Biochemistry},
  volume={11},
  number={5},
  pages={864--878},
  year={1972},
  publisher={ACS Publications}
}

@article{Symmetry-Thiru,
  title={Symmetry, Rigidity, and Allosteric Signaling: From Monomeric Proteins to Molecular Machines},
  author={ Thirumalai,D. and   Hyeon, C and   Zhuravlev, P. I. and  Lorimer G.H. },
  journal={Chemical Reviews},
  volume={119},
  number={12},
  pages={6788--6821},
  year={2019},
  publisher={ACS Publications}
}

@article{horowitz2017minimum,
  title={Minimum energetic cost to maintain a target nonequilibrium state},
  author={Horowitz, Jordan M and Zhou, Kevin and England, Jeremy L},
  journal={Physical Review E},
  volume={95},
  number={4},
  pages={042102},
  year={2017},
  publisher={APS}
}

@article{marko1995stretching,
  title={Stretching dna},
  author={Marko, John F and Siggia, Eric D},
  journal={Macromolecules},
  volume={28},
  number={26},
  pages={8759--8770},
  year={1995},
  publisher={ACS Publications}
}

@article{rief1997reversible,
  title={Reversible unfolding of individual titin immunoglobulin domains by AFM},
  author={Rief, Matthias and Gautel, Mathias and Oesterhelt, Filipp and Fernandez, Julio M and Gaub, Hermann E},
  journal={science},
  volume={276},
  number={5315},
  pages={1109--1112},
  year={1997},
  publisher={American Association for the Advancement of Science}
}

@article{kawakami2006viscoelastic,
  title={Viscoelastic study of the mechanical unfolding of a protein by AFM},
  author={Kawakami, Masaru and Byrne, Katherine and Brockwell, David J and Radford, Sheena E and Smith, D Alastair},
  journal={Biophysical journal},
  volume={91},
  number={2},
  pages={L16--L18},
  year={2006},
  publisher={Elsevier}
}

@article{marszalek1999mechanical,
  title={Mechanical unfolding intermediates in titin modules},
  author={Marszalek, Piotr E and Lu, Hui and Li, Hongbin and Carrion-Vazquez, Mariano and Oberhauser, Andres F and Schulten, Klaus and Fernandez, Julio M},
  journal={Nature},
  volume={402},
  number={6757},
  pages={100--103},
  year={1999},
  publisher={Nature Publishing Group}
}

@article{smith2009fiber,
  title={A fiber-optic interferometer with subpicometer resolution for dc and low-frequency displacement measurement},
  author={Smith, Douglas T and Pratt, Jon R and Howard, LP},
  journal={Review of Scientific Instruments},
  volume={80},
  number={3},
  pages={035105},
  year={2009},
  publisher={American Institute of Physics}
}
\section*{\huge{Supplementary Information}}
\tableofcontents
\setcounter{figure}{0}
\section{Model: Protein - Cantilever System}
There are two aspects to the experiment. Firstly, as the  separation $z$  between the cantilever base and the substrate is increased, force $F$ builds up on the protein that is attached between the cantilever tip and the surface. Eventually, one of the protein domains unfolds due to the force.  The other aspect  involves dynamic measurements where the base of the cantilever is oscillated with a constant frequency and amplitude.  The substrate  retraction is halted at each separation $z$ for a period of 30 ms. During this period the dynamic response of the protein under force $F$ (due to the cantilever base and surface separation $z$) is obtained by recording the amplitude and phase  of the cantilever tip. 
\newcommand\ddfrac[2]{\frac{\displaystyle #1}{\displaystyle #2}}
The equation of motion for the cantilever is given by
\begin{equation*}
    m\Ddot{z} + (k_c+\bar{k})z + \bar{\gamma}\Dot{z} =k_c A_0 \exp{i\omega t} 
\end{equation*}
\\
Here, $z$ is the deflection of the cantilever, $m$ is the effective mass of the moving cantilever, $\bar{\gamma}$ is  the friction coefficient for total damping experienced by the cantilever, $k_c$ is the cantilever stiffness, $A_0$ is the drive amplitude. The $\bar{k}$ and $\bar{\gamma}$ are obtained after linearizing the conservative and dissipative components of the interaction. This assumption is valid because over the oscillation cycle of $1$ nm the interaction force can be considered to be varying linearly.
This is the equation of a linear forced oscillator with the
well-known solutions.

\begin{equation*}
    |A| = \ddfrac{k_c A_0}{\sqrt{(k_c+\bar{k})^2 \Big(1-\frac{\omega^2}{\omega_0^2}\Big)^2 + (\bar{\gamma} \omega)^2}}
\end{equation*}
\begin{equation*}
    \tan{\theta} =- \ddfrac{\bar{\gamma} \omega}{(k_c+\bar{k}) \Big(1-\frac{\omega^2}{\omega_0^2}\Big)}
\end{equation*}
Where $\omega_0$ is given by $\sqrt\frac{k_c+k_i}{m}$. For off-resonance regime the $1-\frac{\omega^2}{\omega_0^2}$ term becomes unity. We can now solve for $\bar{k}$ and $\bar{\gamma}$ to get

\begin{equation*}
    \Bar{k} = k_c\Big(\frac{A_0}{|A|}\cos{\theta}-1\Big)\newline
\end{equation*}
\begin{equation*}
    \Bar{\gamma} = \frac{k_c A_0}{|A|\omega}\sin{\theta}
\end{equation*}
\\
When operating strictly at off-resonance frequency and no molecule is attached to the tip, the phase difference between the cantilever and the drive  is close to zero. The base amplitude and the tip amplitude are same. Hence we can neglect the inertial and velocity dependent forces and the response of the cantilever can be described by its static response.
One can also arrive at the same expression by following a much simpler and intuitive method. For the off-resonance operation, the cantilever and the interaction are in series with each other. The conservative and dissipative components of the interaction are modelled as spring and dash-pot respectively. For such an arrangement we can write the force balance equation as

\begin{equation}
    k_c (A_0 e^{i \omega t} - |A| e^{i( \omega t - \theta)}) = (\bar{k}+i\bar{\gamma}\omega)|A| e^{i(\omega t - \theta)}
\end{equation}
Here $A_0$ and A are the base (drive) amplitude and the tip amplitude respectively. $k_c$ is the cantilever stiffness and $\omega$ is the drive frequency. $\bar{k}$ and $\bar{\gamma}$ are stiffness and friction coefficient of the material beneath the tip. And $\theta$ is the phase lag between the tip and the base.

\begin{equation*}
    k_c (A_0  - |A| e^{-i\theta}) = (\bar{k}+i\bar{\gamma}\omega)|A| e^{-i\theta}
\end{equation*}
\begin{equation*}
    k_c (A_0e^{i\theta}  - |A| ) = (\bar{k}+i\bar{\gamma}\omega)|A| 
\end{equation*}
\begin{equation*}
    k_cA_0 \cos{\theta}+ i k_c\sin{\theta}  - k_c|A| = \bar{k}|A|+i|A|\bar{\gamma}\omega 
\end{equation*}

comparing real and imaginary components gives us
\begin{equation}
    \Bar{k} = k_c\Big(\frac{A_0}{|A|}\cos{\theta}-1\Big)\newline
\end{equation}
\begin{equation}
    \Bar{\gamma} = \frac{k_c A_0}{|A|\omega}\sin{\theta}
\end{equation}

 The stiffness and friction coefficient of the material can then be quantified using equations 2 and 3. To operate at off-resonance, stiff cantilevers,  more than an order of magnitude higher than what are typically used in constant velocity experiments are required. \newline
\begin{figure}[ht]
\centering
\includegraphics[width=\columnwidth]{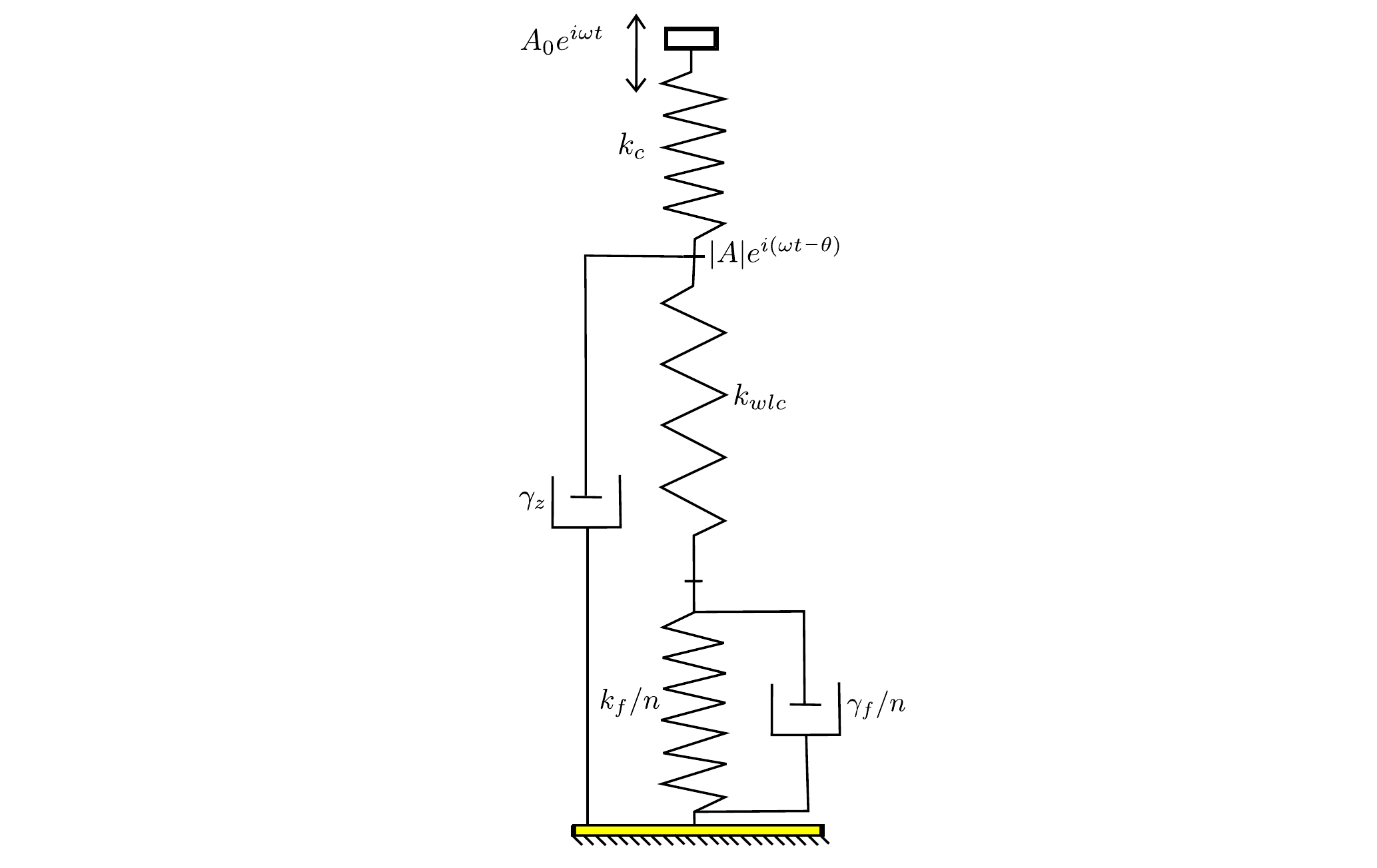}
\renewcommand{\thefigure}{S\arabic{figure}}
\caption{\label{fig:S1}A spring-dashpot  model for the protein-cantilever system. $A_o$ and $A$ are the base and tip amplitude respectively and $\theta$ is the phase lag between them. $\omega$ is the drive frequency. $k_{wlc}$ and $\gamma_z$ are the unfolded chain stiffness and surface damping respectively. $k_f$ and $\gamma_f$ are the stiffness and friction coefficient of the folded domain, where $k_f/n$ and $\gamma_f/n$ means $n$ of them are in series.}
\end{figure}
In the experiment, cantilever, folded domains and the unfolded chain all are in series as shown in Fig S1. The damping provided to the oscillating tip due to proximity of the surface is in parallel to the protein system as shown in the Fig. S1.
For a particular mean tip-sample separation  $z$, the force on all elements such as unfolded chain, folded domains and the linker connecting the domains is same.

\begin{equation}
    k_c(A_0-A)= \Bigg[\Big(\frac{n}{k_{f} + i\gamma_f \omega} + \frac{1}{k_{wlc}}\Big)^{-1} + i\gamma_z\omega\Bigg]A
\end{equation}
Where $n$ is  number of folded domains. 

Rearranging equation 4

\begin{equation*}
    k_c\Big(\frac{A_0}{A} - 1\Big)= \Big( \frac{1}{k_f+i \gamma_f \omega} + \frac{1}{k_{wlc}}\Big)^{-1} + i\gamma_z\omega
\end{equation*}

\begin{equation*}
    k_c\Big(\frac{A_0}{|A|}e^{-i\theta} - 1 \Big) =\Big( \frac{1}{k_{wlc}} + \frac{n}{k_f+i \gamma_f \omega}\Big)^{-1}  + i\gamma_z\omega
\end{equation*}

\begin{equation*}
   k_c\Big(\frac{A_0}{|A|}\cos{\theta} - 1 \Big)+i k_c\frac{A_0}{|A|}\sin{\theta} = \Big(\frac{1}{k_{wlc}} + \frac{n}{k_f+i \gamma_f \omega}\Big)^{-1} + i \gamma_z \omega
\end{equation*}

Using equations 2 and 3
\begin{equation}
    \Bar{k}+i \Bar{\gamma} \omega = \Big(\frac{1}{k_{wlc}} + \frac{n}{k_f+i \gamma_f \omega}\Big)^{-1} + i \gamma_z \omega
\end{equation}
\begin{equation}
     = \Big(\alpha - i \beta \omega)^{-1}  + i \gamma_z \omega
\end{equation}
\begin{align*}
   \text{where}\hspace{0.3cm} \alpha = \frac{1}{k_{wlc}} + \frac{n k_f}{k_f^2 + \omega^2 \gamma_f^2} \hspace{0.3cm} \text{and} \hspace{0.3cm} \beta = \frac{n \gamma_f}{k_f^2 + \omega^2 \gamma_f^2}
\end{align*}
\begin{equation}
    \bar{k}+i\bar{\gamma}\omega=\frac{\alpha}{\alpha^2+\omega^2\beta^2} + i\omega\Big(\frac{\beta}{\alpha^2+\omega^2\beta^2} + \gamma_z\Big)
\end{equation}
The real and imaginary parts of equation can be separated to obtain measured stiffness and friction coefficient  in terms of the model parameters.
\begin{equation}
    \bar{k}=\frac{\alpha}{\alpha^2+\omega^2\beta^2}
\end{equation}
\begin{equation}
    \bar{\gamma}=\frac{\beta}{\alpha^2+\omega^2\beta^2} + \gamma_z
\end{equation}

\subsubsection*{Case 1:} When no protein is attached to the cantilever the $\alpha$ and $\beta$ terms in the equation 8 and equation 9 are $0$. 
This implies that the measured stiffness, in this case, should be zero and the only damping measured should be that provided by  the surface. This is seen when the cantilever is pulled off without a molecule attached to it. See figure S2(b). 
\begin{align*}
    \bar{k} = 0 \hspace{3cm} \bar{\gamma}=\gamma_z
\end{align*}
 --
\subsubsection*{Case 2:}When an octamer is picked up by the tip and all its 8 domains are pulled at, the number of folded domains n = 8. As domains are unfolded sequentially, "n" reduces by 1 at each unfolding event.  However, if the stiffness of folded native states is much higher than the chain stiffness ( $k_{wlc} << k_{f}$), the  second term in $\alpha$ is negligible. Similarly, $\beta$ also is negligibly small.

\begin{align*}
    \bar{k} = k_{wlc} \hspace{3cm} \bar{\gamma}=\gamma_z
\end{align*}
This is also supported by the experiments where the stiffness fits well to WLC model of elasticity as seen in Fig.2 in the main text.
\subsubsection*{Case 3:}
It is known that in the case of I27, when it is pulled with  a force above $\sim$ 90-100 pN, the domain makes a transition to intermediate which is softer as is evident from the data in Fig. 2 in the main text.  

When $n$ domains are in intermediate, measured stiffness and friction coefficient  are given by equation 8 and equation 9 with the respective values of $\alpha$ and $\beta$. 
\begin{align}
  \gamma = \bar{\gamma}-\gamma_z=\ddfrac{\frac{n\gamma_f}{k_f^2 + \omega^2 \gamma_f^2}}{\Big(\frac{1}{k_{wlc}}+\frac{n k_f}{k_f^2 + \omega^2 \gamma_f^2}\Big)^2 + \Big(\frac{n\gamma_f}{k_f^2 + \omega^2 \gamma_f^2}\Big)^2}
\end{align}

\begin{align}
    \bar{k}=\ddfrac{\frac{1}{k_{wlc}} + n\frac{k_f}{k_f^2+\omega^2\gamma_f^2}}{\Big(\frac{1}{k_{wlc}} + n\frac{k_f}{k_f^2+\omega^2\gamma_f^2}\Big)^2 + \Big(\frac{n\gamma_f}{k_f^2 + \omega^2 \gamma_f^2}\Big)^2}
\end{align}

 Where $\gamma = \bar{\gamma} - \gamma_z$ is damping provided by the protein alone. It is obtained by subtracting out the background due to surface damping by fitting a polynomial. See Fig. S2(a).   
\begin{figure}[ht]
    \centering
    \includegraphics[width=\columnwidth]{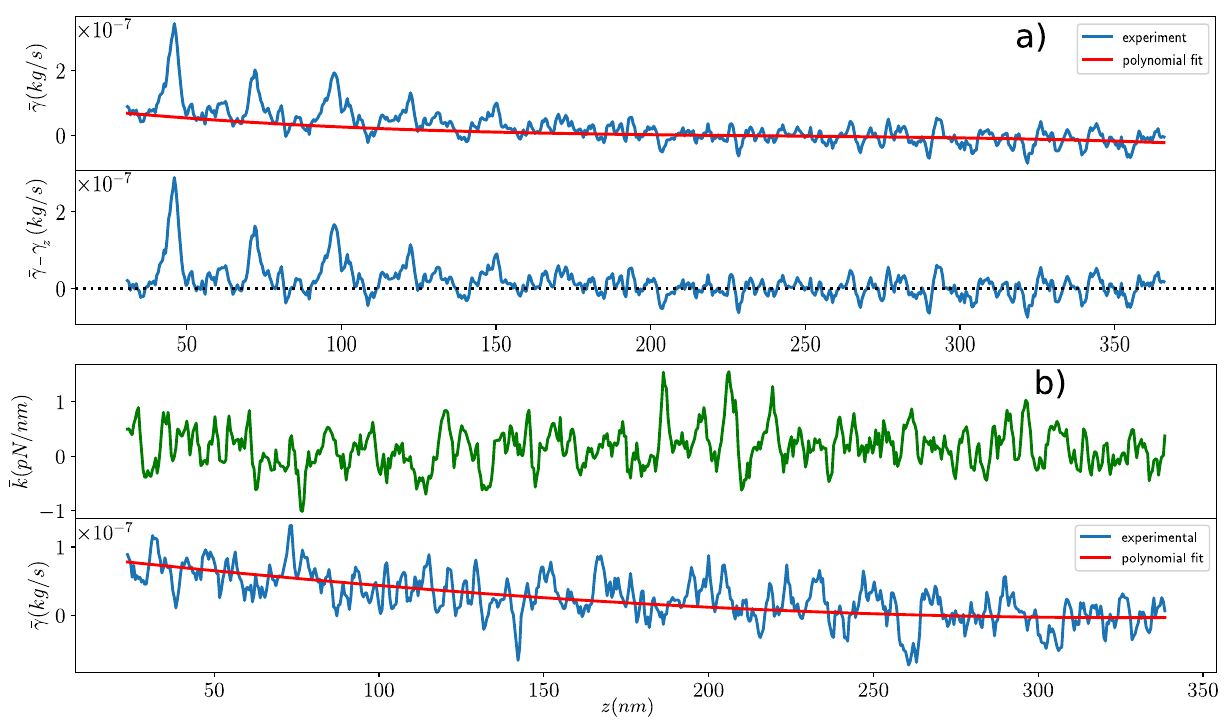}
    \renewcommand{\thefigure}{S\arabic{figure}}
    \caption{a)  The measured total friction  coefficient (blue) along with a polynomial fit (red) for the damping due to the surface effects. The lower panel in a) represents friction  coefficient of the protein alone,  $\bar{\gamma}-\gamma_z$
    b) shows that measured stiffness is zero   when the protein is not attached, however the damping due to the surface is still measured (blue).}
  \label{fig:S2}
\end{figure}  
 
\subsubsection*{Case 4:}
With each unfolding event, the unfolded chain's contour length increases by  29 nm. The larger contour length makes the chain more complaint.  The unfolding force is reached before the chain is extended to have a stiffness which is comparable to the folded intermediates (see Fig S8). In this region usually after four unfolding events, the folded intermediate's stiffness again becomes much larger compared to  the unfolded chain $k_{wlc}<<k_f$. The surface damping at such large separations can also be neglected as is seen in the data. This reduces equation 10 and 11 to
\begin{align*}
    \bar{k} = k_{wlc} \hspace{3cm} \bar{\gamma}=\gamma_z=0
\end{align*}
\FloatBarrier

Since our  model and its special cases describe features observed in our  experiments, we  use equation 4 to get stiffness and friction coefficient  of folded intermediates.
\begin{equation}
    \Big(\frac{1}{\bar{k}+i(\bar{\gamma}-\gamma_z)\omega} - \frac{1}{k_{wlc}}\Big)^{-1} = \frac{k_f}{n} + i\frac{\gamma_f \omega}{n}
\end{equation}
Simplifying this equation leads to the expression for stiffness and friction coefficient  of the folded domains in terms of the experimentally measured quantities.
\begin{equation}
       k_f=\ddfrac{n\Big(\frac{\bar{k}}{\bar{k}^2 + \omega^2 \gamma^2} - \frac{1}{k_{wlc}}\Big)}{\Big(\frac{\bar{k}}{\bar{k}^2 + \omega^2 \gamma^2} - \frac{1}{k_{wlc}}\Big)^2 + \Big(\frac{\gamma\omega}{\bar{k}^2 + \omega^2 \gamma^2}\Big)^2} 
\end{equation}
\begin{equation}
       \gamma_f=\ddfrac{n\Big(\frac{\gamma\omega}{\bar{k}^2 + \omega^2 \gamma^2}\Big)}{\Big(\frac{\bar{k}}{\bar{k}^2 + \omega^2 \gamma^2} - \frac{1}{k_{wlc}}\Big)^2 + \Big(\frac{\gamma\omega}{\bar{k}^2 + \omega^2 \gamma^2}\Big)^2} 
\end{equation}
Where all the quantities on the right hand side are known experimentally. $k_{wlc}$ is obtained by fitting wlc to the  stiffness-extension data. In the next section, we describe our methodology to obtain $k_{wlc}$ from our experimental data. Using equations 13 and 14, the stiffness and friction  coefficient of folded intermediates are obtained when all the folded domains n are in the intermediate state. In experiments, this is marked by the dark shaded region in Fig. 2 of the main text, where the measured stiffness starts to rise again.  
\newpage
\section{Data Analysis}

\subsection{Analysis  to obtain \texorpdfstring{$k_{wlc}$}{kf}}
\textit{Fitting the wlc model:}
In the initial part of rising stiffness (region in Fig. 2, not shaded,  in the manuscript) for each unfolding event, all the folded domains are in the native state,  where $ (k_f >> k_{wlc})$. 
This leads to equation 8 reducing to $\bar{k} = k_{wlc}$, as discussed in the previous section. The contribution to  measured stiffness in this  region is entirely from the unfolded chain, as is evident from the wlc model fits.
The differential form of the wlc model \autocite{marko1995stretching} which relates $k_{wlc}$ to its end to end distance is fitted to the measured stiffness. The contour length $L_c$ is the free parameter and the persistence length $p$ is taken as $0.4$ nm \autocite{rief1997reversible,kawakami2006viscoelastic}.

\begin{equation}
    \frac{dF}{dz}=\frac{k_BT}{pL_c}\Big(\frac{1}{2(1-z/L_c)^3}+1\Big)
\end{equation}

It is observed that the wlc fits to the experimental data up to a certain value of the extension $z$, as seen in Fig. 2.

\textit{ Reconstructing  force profiles from   stiffness data:} In order to estimate the contribution from the unfolded chain, $k_{wlc}$, to the total measured stiffness   (shaded region in Fig. 2 manuscript), we need to evaluate force in this region. This is obtained by integrating the stiffness-extension data. This integration is done in two parts. 
First, this integration is done over the fitted wlc curve. Second, the numerical integration is carried out on the experimental data in the  region where  the wlc deviates from the measured  stiffness-extension profile (shaded in blue ).      
After obtaining the force profile in the shaded blue region, one can use the following  to estimate $k_{wlc}$. The relationship is shown to work in the region of high stretch \autocite{khatri2008internal}.
\begin{equation}
    k_{wlc}(F)=\frac{4}{L_c}\sqrt{\frac{p}{k_BT}}F^{\frac{3}{2}}
\end{equation}
\FloatBarrier
\begin{figure}[ht]
\centering
\includegraphics[width=\columnwidth]{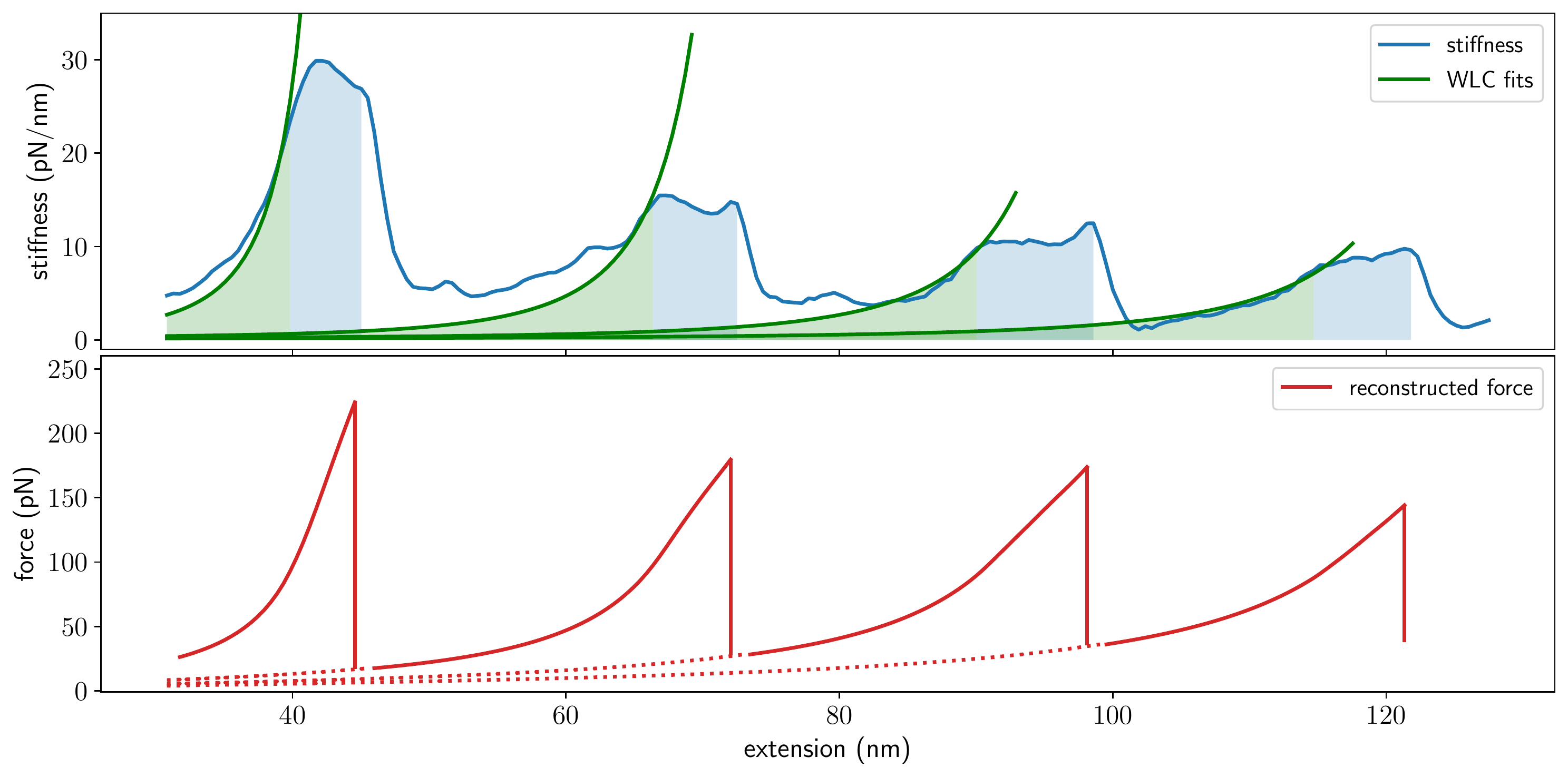}
\renewcommand{\thefigure}{S\arabic{figure}}
\caption{\label{fig:S3}Numerical integration of the stiffness-extension  profiles to obtain force-extension profiles. Continuous green curve- wlc fits, Continuous blue curve- experimental data, continuous red curve -  force obtained after integration. The green shaded region represents integration done over fitted WLC model. The blue shaded region represents integration over experimental data where it deviates from WLC behavior.}
\end{figure}
\FloatBarrier

\newpage
\subsection{Obtaining \texorpdfstring{$k_f$}{kf} and \texorpdfstring{$\gamma_f$}{gammaf}}In equation 16, the contour length $L_c$ needs to be evaluated for the analysis of the shaded region. Note that in this region, some of the folded domains are in the native state and some are in intermediate. The intermediate is formed by breaking hydrogen bonds between A and B $\beta$-strands. Once broken, the  $\beta$-strand A  is released adding 0.66 nm to the contour length\autocite{marszalek1999mechanical,nunes2010force,taniguchi2008effect}. As a result, it is difficult to obtain the exact value of $L_c$ in the shaded region of a given unfolding event. Secondly, the number of domains  in the  intermediate $n$,   are needed in equation 13 and 14 to obtain the stiffness of folded intermediate. One of the noteworthy features of our data is that towards the end of the shaded region the measured stiffness starts to rise again. We argue that at this point, all the  folded domains are in the intermediate. This allows us to use $n$ in equation 13 and 14 and $L_c$ in equation 15 becomes $L_c + n0.66$. With this, the $k_{wlc}$ at each force $F(z)$ is determined and equation 13 and 14 are then used to estimate the stiffness and friction  coefficient  of folded intermediate. Note that we are able to measure the stiffness of the folded intermediate using this methodology, the stiffness of the native folded state is immeasurably high.

The analysis scheme to obtain $k_f$ and $\gamma_f$ is repeated over many data sets. These data sets are shown in Fig. S4-S7. The grey shaded region is used for the said data analysis.  The values obtained at different forces are then plotted together in figure 4c and d in the main text.

\begin{figure}[ht]
\centering
\includegraphics[width=0.9\columnwidth]{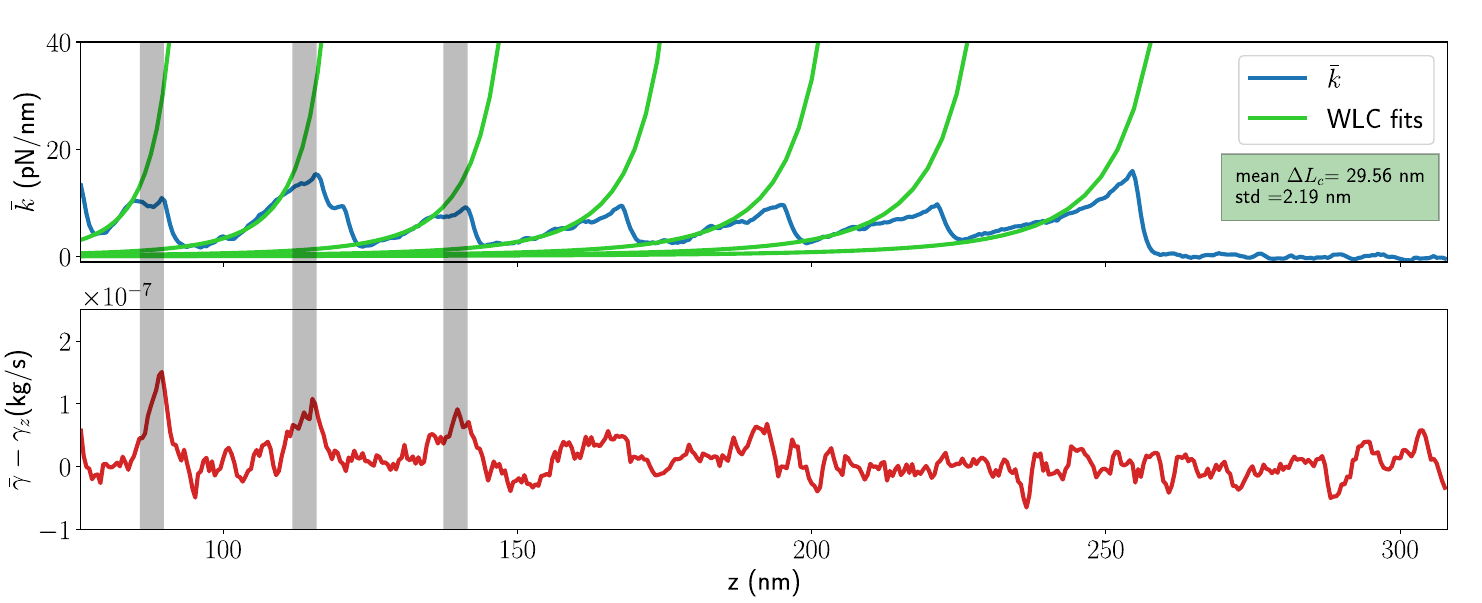}
\renewcommand{\thefigure}{S\arabic{figure}}
\caption{\label{fig:S4} }
\end{figure}

\begin{figure}[ht]
\centering
\includegraphics[width=0.9\columnwidth]{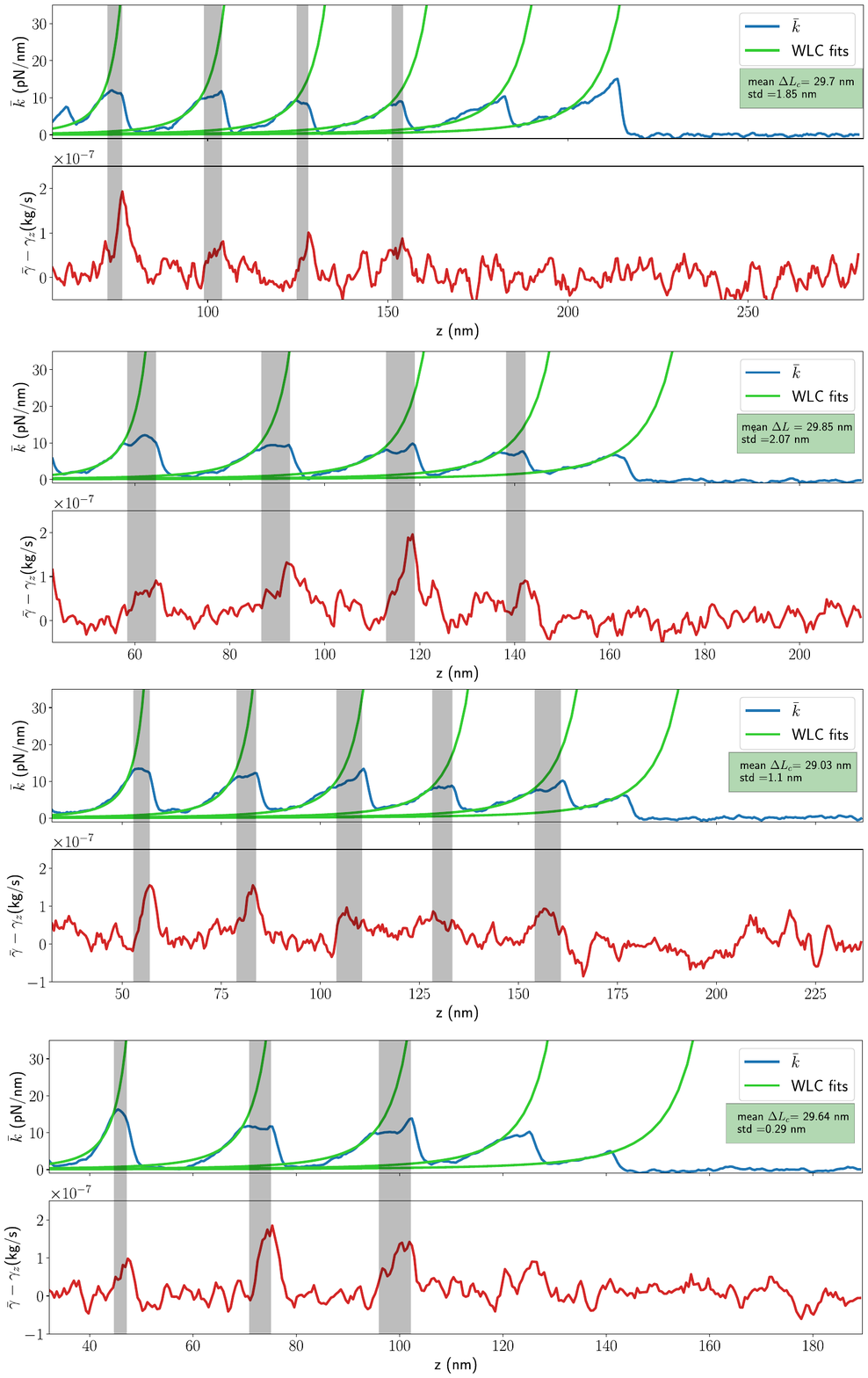}
\renewcommand{\thefigure}{S\arabic{figure}}
\caption{\label{fig:S5} }
\end{figure}

\begin{figure}[ht]
\centering
\includegraphics[width=0.9\columnwidth]{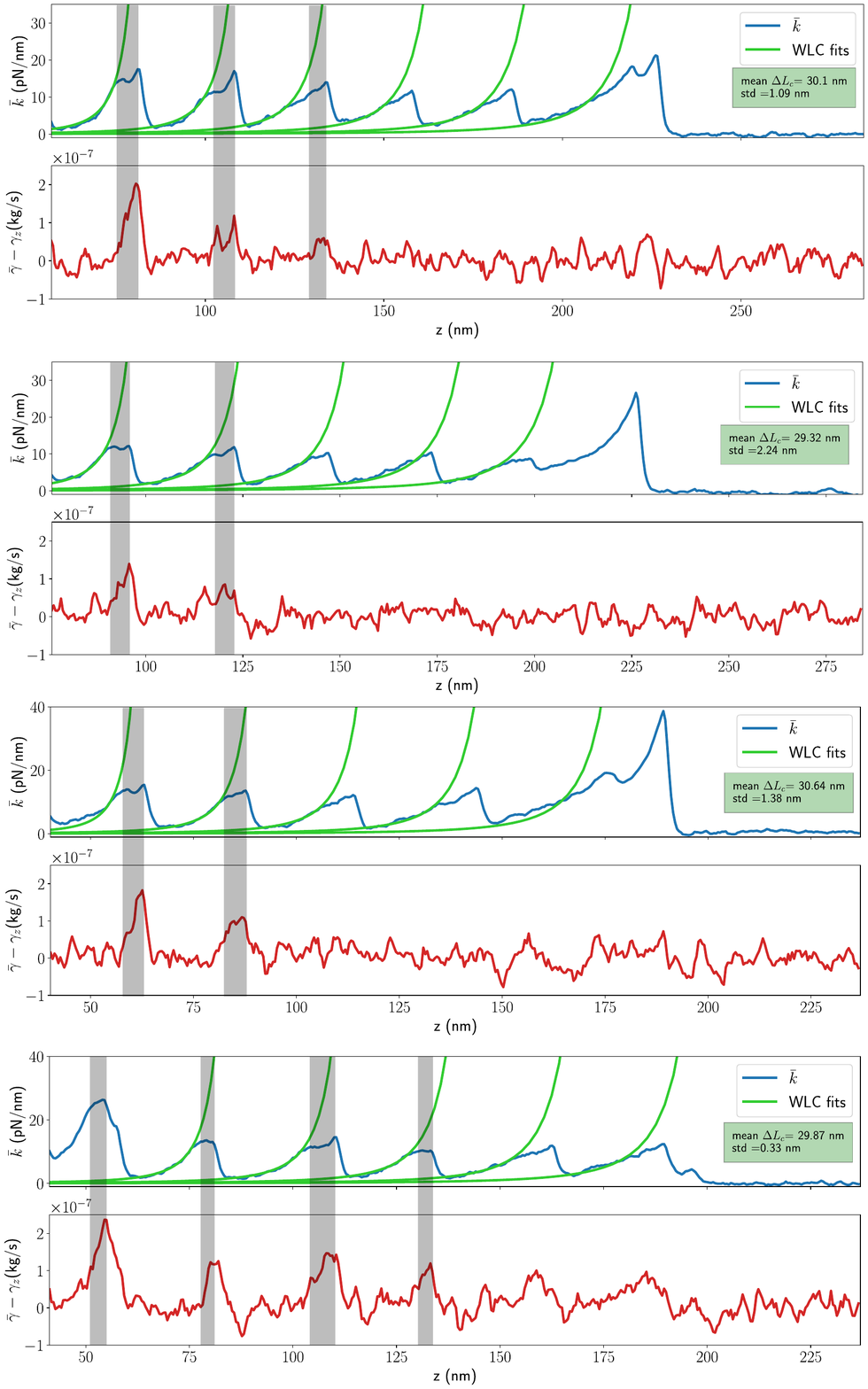}
\renewcommand{\thefigure}{S\arabic{figure}}
\caption{\label{fig:S6} }
\end{figure}

\begin{figure}[ht]
\centering
\includegraphics[width=0.9\columnwidth]{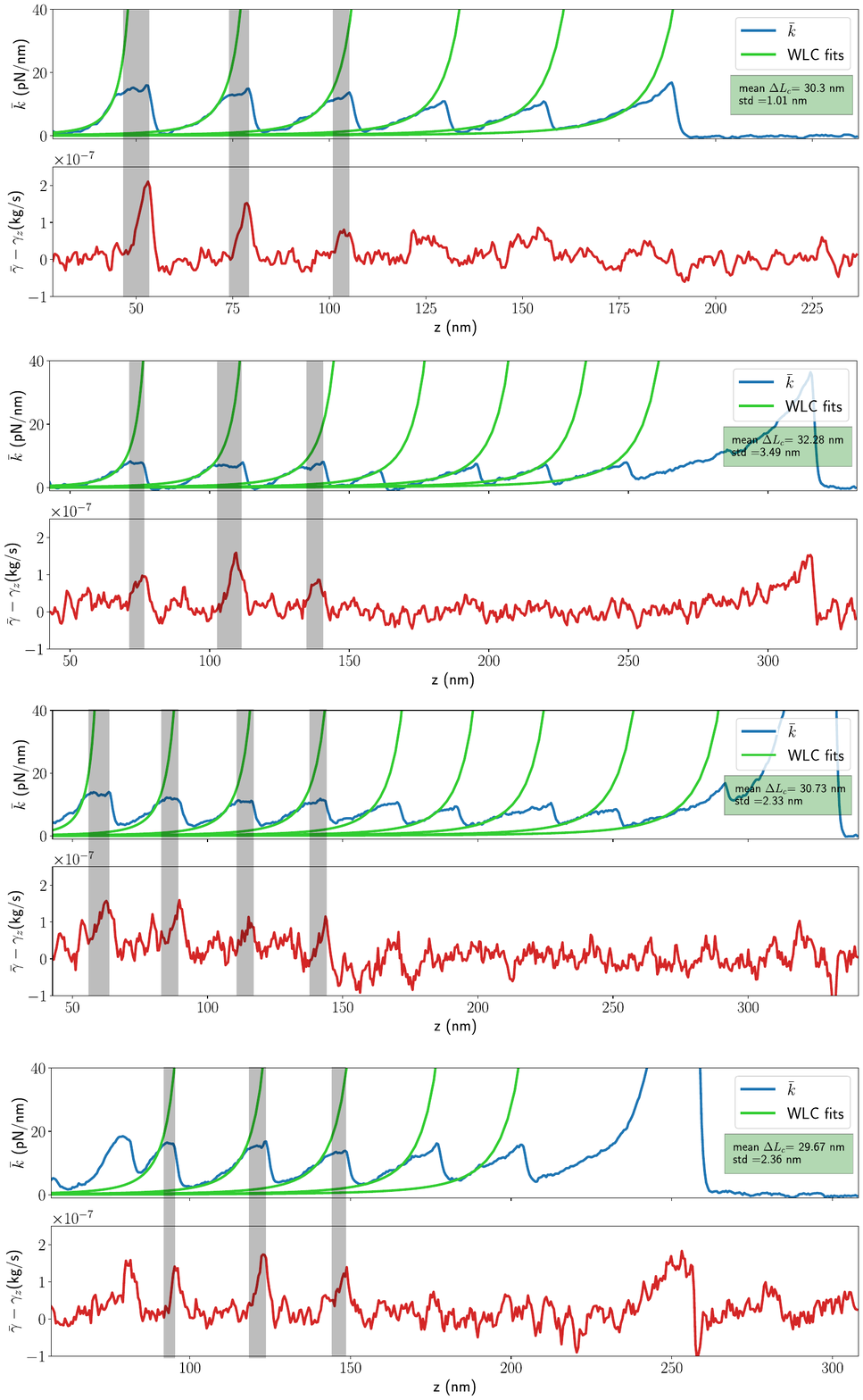}
\renewcommand{\thefigure}{S\arabic{figure}}
\caption{\label{fig:S7} The total stiffness-extension  and total friction coefficient-extension profiles. The  analysis of these curves yields stiffness and  friction coefficient of individual domains,  plotted in Fig. 4c and d.}    
\end{figure}
\FloatBarrier
\section{Features observed in stiffness-extension and friction coefficient-extension profiles}    
\subsection{The wlc plots of stiffness verses extension and Force verses extension}
The force-extension behavior of a polymer chain is described by Marko-Siggia approximation of Worm-Like-Chain (wlc) model\autocite{marko1995stretching}. According to the model,  which has been shown to work for experimental  data obtained using AFM and optical tweezers, the force versus extension is given by

\begin{equation}
F= \frac{k_BT}{p}\Big(\frac{1}{4{(1-z/L_c)}^2} - \frac{z}{L_c}- \frac{1}{4}\Big )
\end{equation}

The derivative $F$  of this with extension $z$ is stiffness $k_{wlc}$ 
\begin{equation}
    k_{wlc}=\frac{dF}{dz}=\frac{k_BT}{pL_c}\Big(\frac{1}{2(1-z/L_c)^3}+1\Big)
\end{equation}
\begin{figure}[ht]
 \label{fig:S8}
\FloatBarrier
  \centering
    \includegraphics[width=\columnwidth]{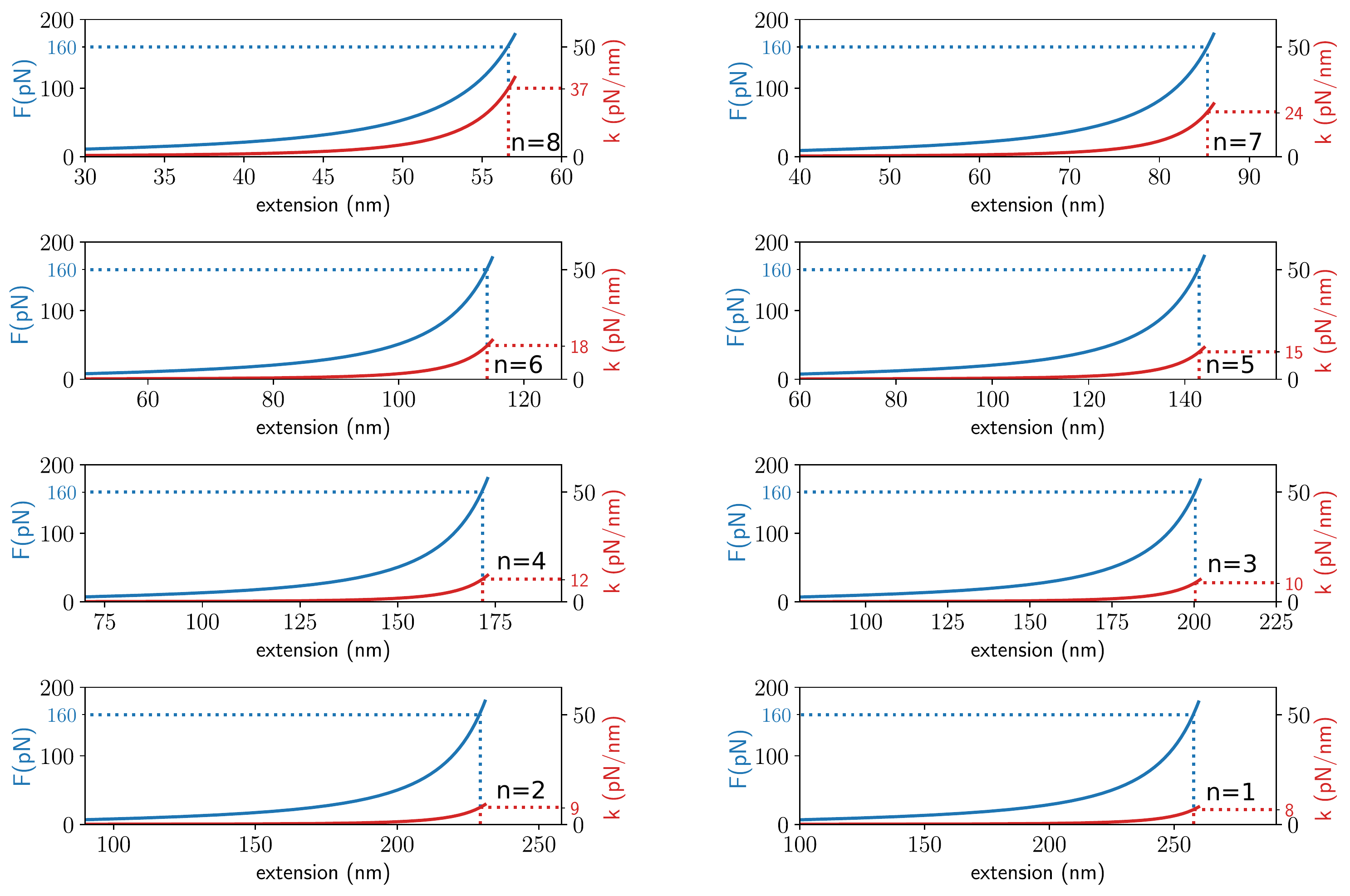}
    \renewcommand{\thefigure}{S\arabic{figure}}
    \caption{Force and stiffness variation for each contour length. With each unfolding event, the unfolded chain becomes more complaint for the same force ($\sim 160$) pN, thereby dominating the total measured stiffness. The protein is unfolded before the chain  has a comparable stiffness     to the folded domains so  that they start to  contribute to the total stiffness measurement. Thus, last four domains unfold without contributing to the stiffness measurement and wlc fits to the entire stretching of the chain.    }
    \end{figure}

Fig. S8 shows plots of these two equations for a fixed persistence length $p$ = 0.4 nm and different contour lengths $L_c$.  The $L_c $ depends on   number of folded domains $n$. As  domains are sequentially unfolded in AFM experiments, each unfolded domain adds 29 nm to the total contour length. The  plots clearly indicate  that the stiffness of the chain rises sharply compared to the force for larger contour lengths. This means that the  chain is more complaint  under a given force for larger contour lengths. When there are 8 folded domains, the stiffness is 37 pN/nm for 160 pN, the force at which the domain unfolds. It is comparable to the domain stiffness.   This progressively decreases to  12 pN/nm when there are only 4 domains which are folded. It means that next domain  unfolds before the chain becomes stiff enough, so that folded domains contribute to the total stiffness measurement. For last four domains, the contribution to the total stiffness by folded domains is immeasurably low and the wlc model fits to the entire stretching of the molecule till next domain unfolds.  It is noteworthy that the dissipation is also close to zero in this region.

\begin{figure}[ht]
 \label{fig:S9}
\FloatBarrier
  \centering
    \includegraphics[width=\columnwidth]{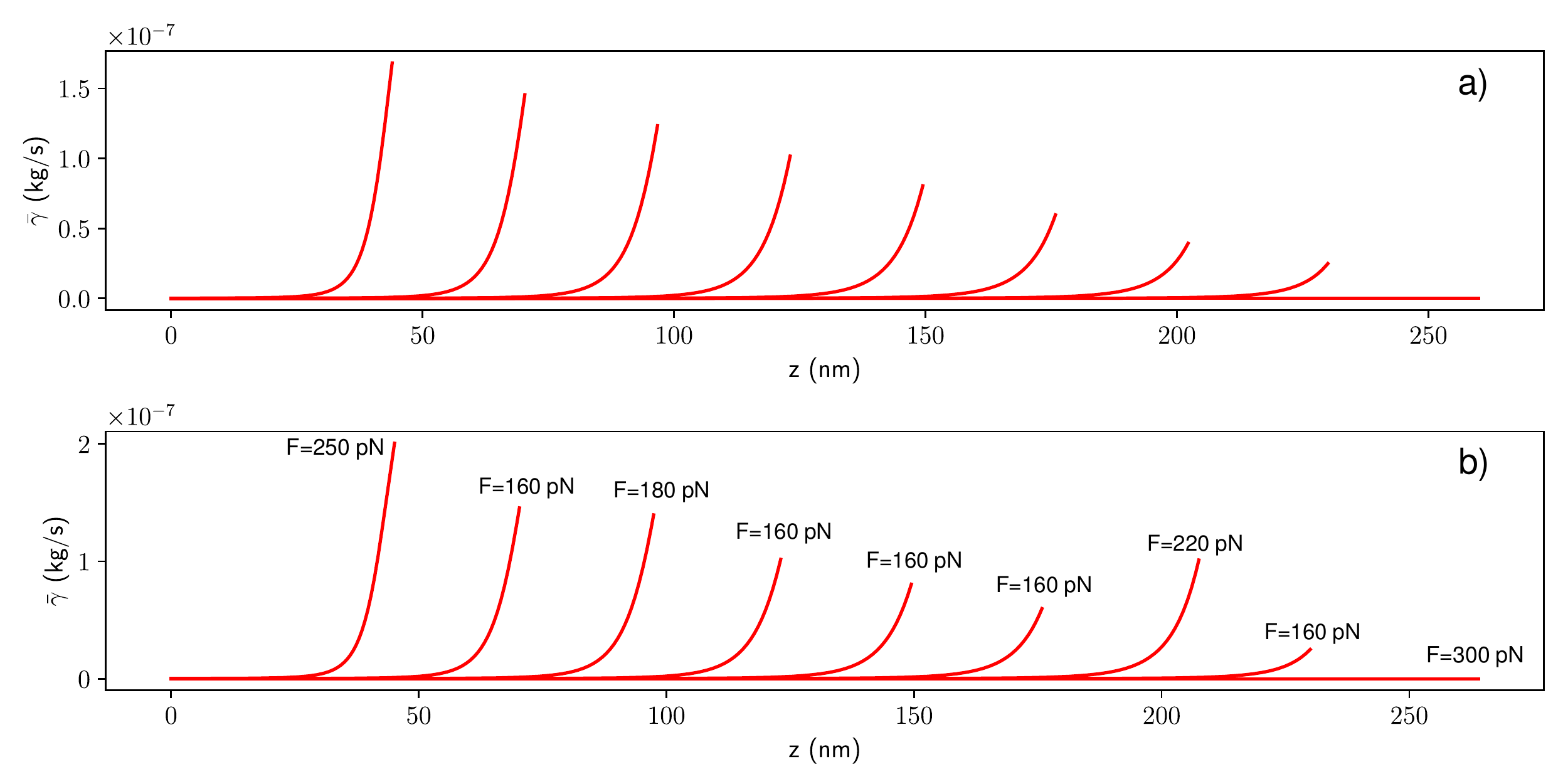}
    \renewcommand{\thefigure}{S\arabic{figure}}
    \caption{a)The plot of total friction coefficient $\bar{\gamma}$  calculated at  different tip-substrate separation $z$ using equation 10. Here the $\gamma_z$ is taken  to be zero.  The contour length is increased by 29 nm  after each event and $n$ is reduced by one. $k_f$ and $\gamma_f$ are taken as 100 pN/nm and $2 \times 10^{-6}$ kg/s. The variation in friction coefficient with respect to $z$ is plotted for each event until 160 pN is reached. This is the average force at which domain unfolds. b) Each domain unfolds at slightly different force as the unfolding events are stochastic.   We take  similar values for all other   parameters as in a, however $\bar{\gamma}$ is now calculated and plotted for each event until unfolding force values are reached for that event.     In both plots,  it is seen that the peak in the total friction coefficient decreases as  domains unfold one by one. The height of the peak also depends on the unfolding force for that event. In experiments, friction coefficient after fourth domain is not observed due to reasons provided in the previous subsection. }
\end{figure}

\subsection{Decreasing peak in total friction coefficient  for  consecutive unfolding events}
 It is observed that the total friction coefficient peak height decreases as more and more domains unfold. This feature also needs explanation.   It should be noted that the measured damping coefficient is not only of the domain that is unfolding. It is a combined response of all the domains and depends on many other quantities.
The $\bar{\gamma }$ given by equation 10,   depends not only on friction coefficient of individual domain $\gamma_f$, but also on   stiffness of the wlc chain $k_{wlc}$, the number of folded domains $n$, and stiffness of the individual domain $k_f$. Of these, we argue that the $\gamma_f$ and $k_f$ remains fixed. Taking them as constants, we plot equation 10 in Fig. S9a to obtain the total damping coefficient $\bar{\gamma}$ as a function tip-substrate separation till 160 pN is reached and reducing $n$ by 1 after each unfolding event and increasing the $L_c$ by 29 nm. The plot clearly shows that the peak $\bar{\gamma}$ is decreasing with  increasing unfolding events. Moreover, we plot the $\bar{\gamma}$ till the unfolding force is reached for the individual  unfolding events, if these events are stochastic and the unfolding force for individual domains  differ from one another.   This  results in non-monotonous decrease in the total $\bar{\gamma}$ for each peak as seen in Fig. S9b. The plots explain the observed decrease in the total friction coefficient as the domains sequentially unfold.

\section{ Fibre-Interferometer based AFM  and  single molecules }
In a typical force spectroscopy experiment, wherein  the cantilever    is oscillated either at the base or the tip-end,  it is treated as a point mass having mass $m$ and spring constant $k$. Typically this works in ambient or ultra-high vacuum conditions, where quality factors are  large (100 -10000). It has been increasingly realized that such modelling will fail in viscous media such as water and other liquids where the quality factor is low (1-5). When experiments are performed on-resonance and the resultant amplitude and phase are used to quantitatively obtain the stiffness and friction coefficient of the system beneath the tip, it is difficult to predict the phase behaviour seen in the experiments by modelling the cantilever alone\autocite{o2006comment,kiracofe2011quantitative}. In particular, for single-molecule experiments, it poses additional challenge of controlling the extensions produced in the molecule in a given oscillation cycle.

The above discussion necessitates the off-resonance operation to measure the viscoelasticity of single molecules. There are attempts in the literature to measure single molecule viscoelasticity by driving the cantilever off-resonance\autocite{benedetti2016can,janovjak2005molecular,kawakami2006viscoelastic,taniguchi2010dynamics}. However, the cantilever stiffness (and resonance frequency ) in all these measurements is low ($\sim 0.020$ pN/nm) and does not allow  true off-resonance conditions. In such situations, for a  driven cantilever- from either base or tip -, the amplitude reaches its maximum value at resonance and  the phase is  approximately 90 $^{\circ}$ (Fig. S10). When the tip  interacts with a purely elastic element,  the resonance of the  system shifts to a higher frequency fig.S10b. If one is not  operating  at \textit{true} off-resonance, such shifts produce a phase signal. If  the molecule has internal friction, it is very difficult to separate it from the effects produced due to stiffness change.
 In experiments especially in a liquid environment, it becomes very difficult to predict how the phase will behave when the interaction contains both conservative and dissipative components.

This  issue is  resolved if  the experiment is performed using stiff cantilevers (0.6 - 1 N/m) having high resonance frequencies  (25 - 50 KHz in water).   This  ensures  true  off-resonance operation where the phase lag is close to zero and any shift in resonance due to stiffness change does not cause  any phase shift.  With these cantilevers we get a window of off-resonance condition, in which  the measured phase lags are  result of   damping provided by the molecule alone. See Fig. S10(c) and (d) 
 In single-molecule measurements, the measured stiffness of the poly-protein system  is of the order of 10 pN/mm. In true off-resonance, this will produce 1 percent change in tip amplitude. Moreover, it will produce an even smaller bending signal. ( For  instance  with a base amplitude of 1 nm, the tip amplitude is 0.99 nm) This produces a bending amplitude ($A_0-A$)  of 10 pm. Such signals can not be measured using a conventional deflection detection scheme, which is equipped to measure the bending alone . See figure S11 and S12.   The interferometer based AFM  measures  tip displacements. This allows  measurement of  the  extension produced in the molecule directly, in contrast to the deflection detection scheme,   which measures bending in the cantilever (Fig S11). 
 
 Secondly,  a free-of -artefacts,   off-resonance operation,   is possible with tip excitation and measuring the bending in the cantilever. Here the complications associated with phase response are absent.  However, the cantilever and the protein system are in parallel with each other now and the stiffness of the protein system is roughly 100 times smaller than the cantilever ( Cantilever: 1 N/m and protein: 10 mN/m). The change in amplitude after protein is attached to the tip is thus 1 part in 100 and is difficult to detect. One needs to use a cantilever having stiffness  comparable to that of protein. This  makes off-resonance operation difficult as shown in Figure S10.    
  Hence, for measuring the viscoelastic response of protein by exciting the cantilever,  interferometer based measurements with base excitation are most suitable. Currently,  the measurement bandwidth is limited due to noise floor at low frequencies. See figure S13. This can be overcome by improving laser coherence and filters \autocite{smith2009fiber}.    

 \FloatBarrier
 \begin{figure}[ht]
\centering
\includegraphics[width=\columnwidth]{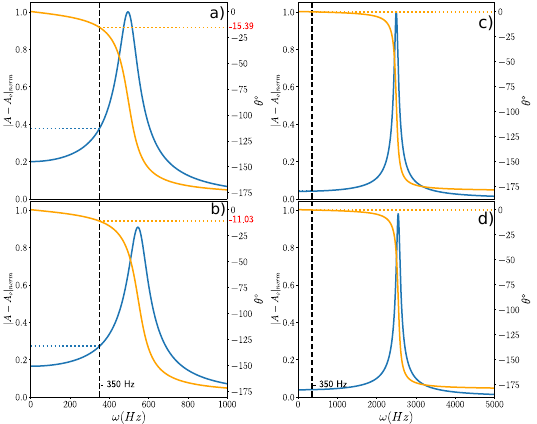}
\renewcommand{\thefigure}{S\arabic{figure}}
\caption{The frequency response of a cantilever which is given by the damped driven harmonic oscillator model: $|A-A_o|=\frac{F_0}{m}\frac{1}{\sqrt{(\omega^2-\omega_o^2)^2+(\gamma\omega)^2}}$ and $\tan \theta = -\frac{\gamma \omega}{\omega_o^2-\omega^2}$ where $F_o$ is the amplitude of the forcing, $|A-A_0|$ represents the bending since $A$ and $A_o$ are the tip and base amplitudes respectively. In the figures  $|A-A_o|_{norm}$ is the normalised amplitude. a) and c) have the same parameters except for the resonance frequency, $\omega_o$ in c) is $10$ times that of a), however the drive frequency is the same for both. In (c) the drive frequency is much lower in comparison to its resonance frequency $\omega_o >> \omega$ and the amplitude response can be approximated by the static response: $A-A_o \approx \frac{F_0}{k_c}$ and $\theta \approx 0$. In b) and d) an interaction is incorporated that increases the resonance frequency in a) and c) by the same amount. As is evident from the figure, in  b) we see a significant change in phase lag, whereas in  d)  very little change is seen.
It is essential to note that one cannot perform off-resonance operation with a cantilever whose response is similar to a), because even at drive frequencies as low as what has been shown here,  there is no flat regime in the phase response, any stiffness change can still bring about a change in phase. It is only when stiff cantilevers with high resonance frequencies are used one can achieve this off-resonance operation.}
\label{fig:S10}
\end{figure}
 \FloatBarrier

\begin{figure}[ht]
\centering
\includegraphics[width=0.9\columnwidth]{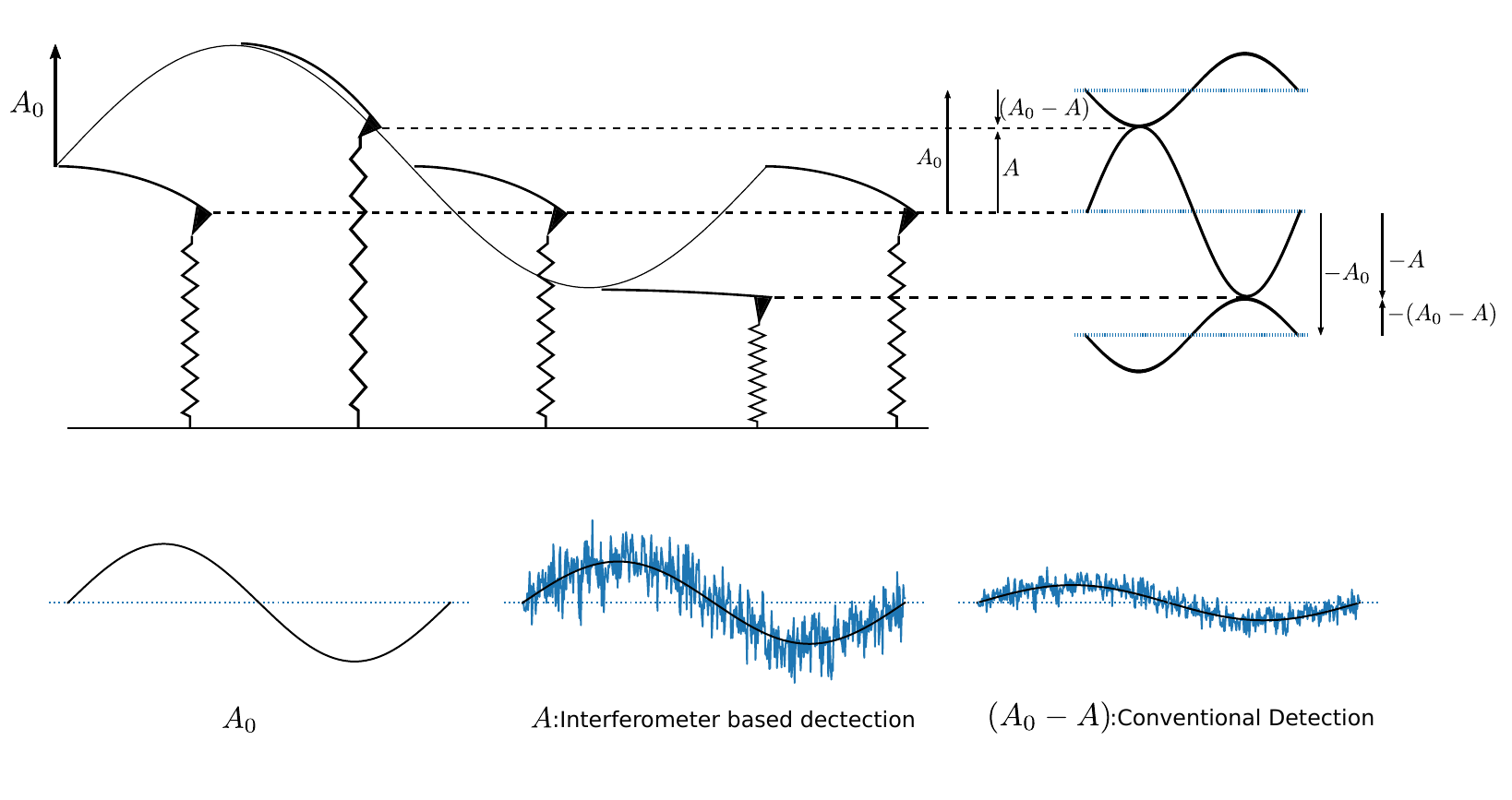}
\renewcommand{\thefigure}{S\arabic{figure}}
\caption{\label{fig:S11} The  difference in tip displacement signal ($A$) and the cantilever bending signal($A_0-A$). Interferometer based detection measures tip displacements whereas the deflection detection scheme measures the bending. For stiff cantilevers(0.5 to 1 N/m), the bending is extremely small and difficult to detect. See fig. S12 for experimental data which shows the actual bending  and displacement signals. 
 Furthermore, $A$ also represents the amplitude of extension in the molecule over the oscillation, which is directly measured by the interferometer.}
\end{figure}

\begin{figure}[ht]
\centering
\includegraphics[width=\columnwidth]{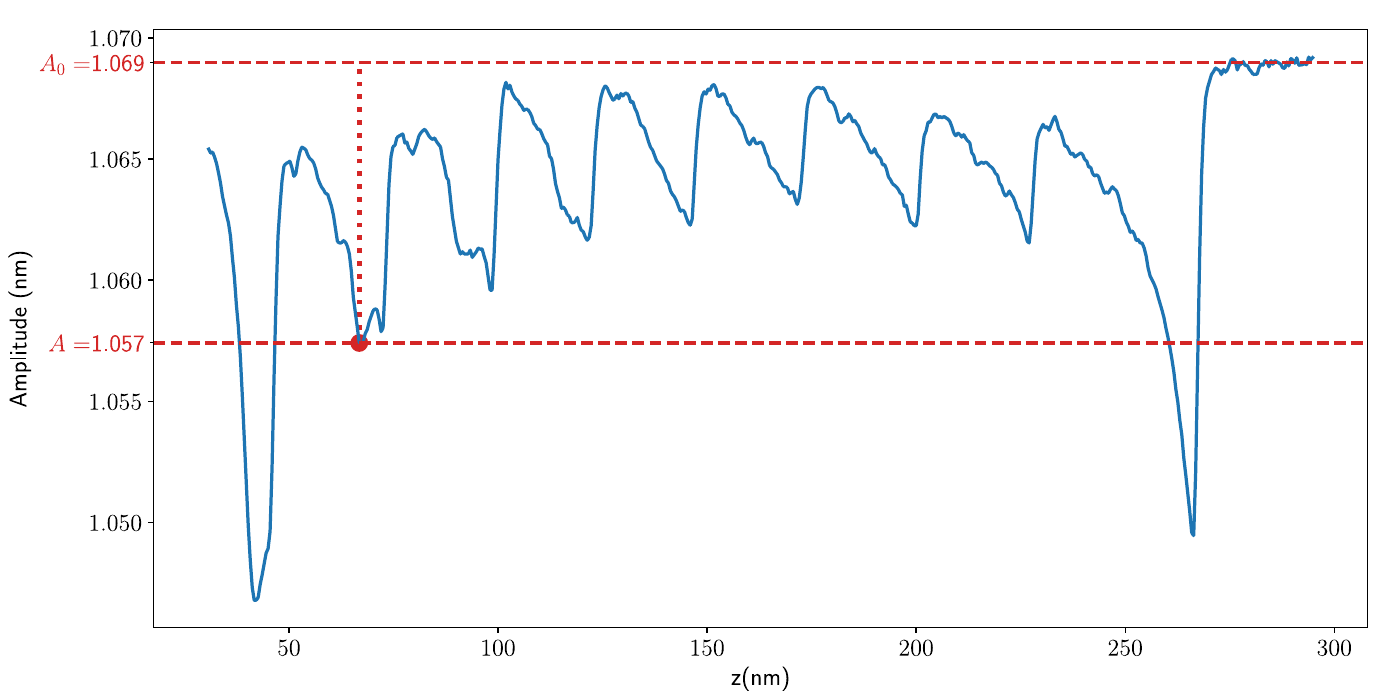}
\renewcommand{\thefigure}{S\arabic{figure}}
\caption{\label{fig:S12}The amplitude $A$ change measured as the cantilever is pulled away from the surface with the protein attached. $A_o$ is determined from the measured amplitude when the protein detaches from the cantilever. The bending signal in this measurement at the point marked in red $A_0-A$ is $12pm$  The sensitivity of the interferometer for is measurement was $300$ mV/nm}
\end{figure}
\FloatBarrier
\section {Measurement at lower frequency}
In order to see the effect of performing measurements at lower frequencies, the experiments are carried out at 433 Hz. The stiffness data once again shows that at high stretch ( shaded region in Fig.S8a ) there is contribution to stiffness from folded domains. The stiffness-extension data begins to deviate from the wlc description. We use our analysis in this region to separate the stiffness of folded intermediates from the combined total stiffness.  The stiffness of folded intermediates is similar to measurement performed at $\sim$ 2 KHz. The friction coefficient signal, on the other hand is featureless. However the noise floor at this frequency is much higher compared to the frequency window around 2 KHz. It can not be concluded from this data if the internal friction coefficient  depends on operational frequency.       
\begin{figure}[ht]
\centering
\includegraphics[width=\columnwidth]{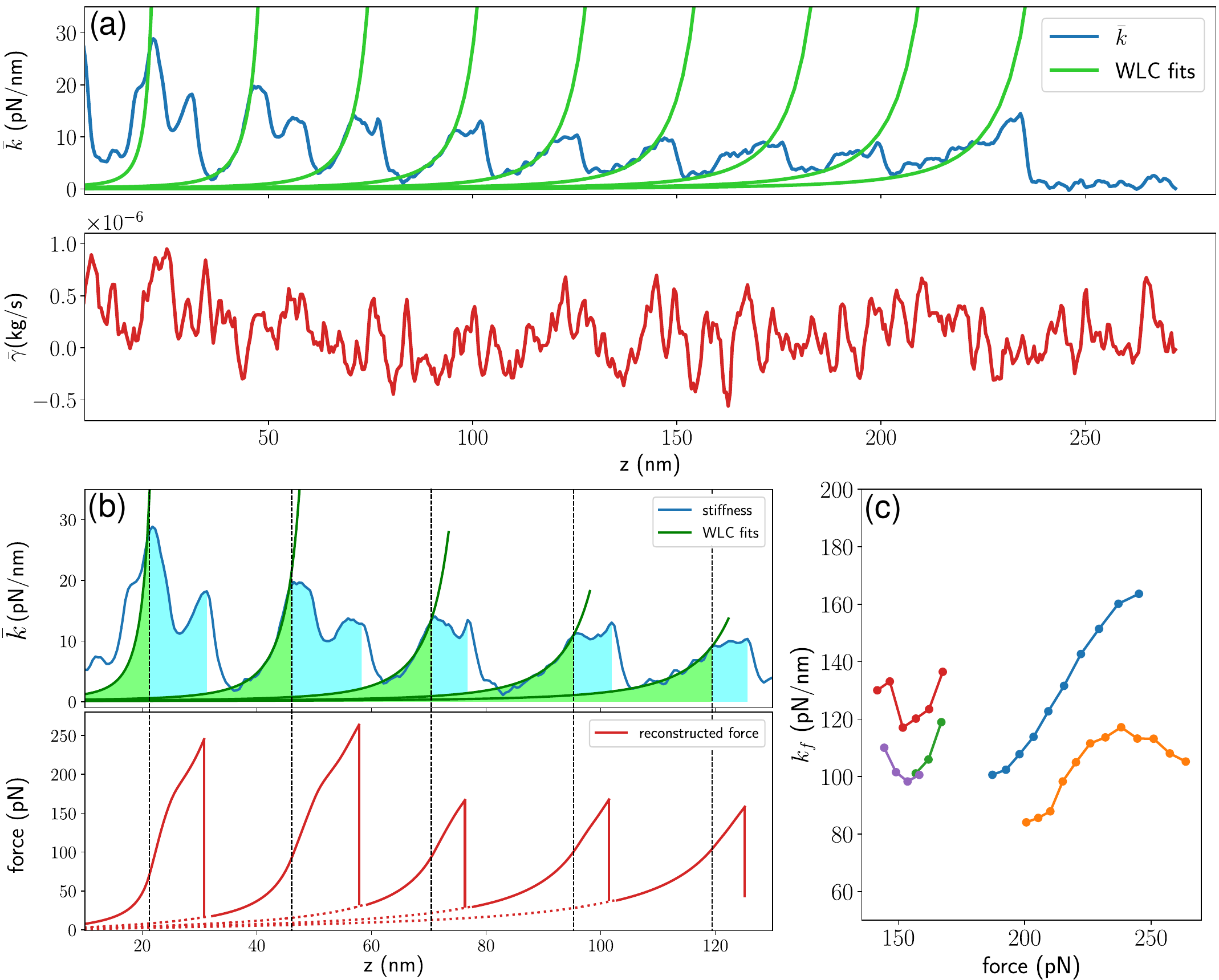}
\renewcommand{\thefigure}{S\arabic{figure}}
\caption{\label{fig:S13}Measured stiffness and friction coefficient for a measurement performed at 433Hz. The noise floor is higher than the measured friction coefficient  at 2.1 kHz. The dissipation is hence not measured. The analysis shows that the force profiles and its intersection with dotted vertical lines  once again indicates that the transition to region, where stiffness from folded intermediates contribute to the measurement occurs at $\sim$ 95 pN. The stiffness of the intermediate at different forces are also shown. The stiffness values are similar to the experiment performed at higher frequencies around 2 KHz.  }
\end{figure}

\end{document}